\documentclass{pnastwo}

\url{www.pnas.org/cgi/doi/10.1073/pnas.0709640104}
\copyrightyear{2008}
\issuedate{Issue Date}
\volume{Volume}
\issuenumber{Issue Number}
\usepackage{bm}

\begin{document}

\title{Nanodomains in biomembranes with recycling}

\author{Mareike Berger\affil{1}{Laboratoire de Physique Th\'eorique, IRSAMC, Universit\'e de Toulouse, CNRS, UPS,
118 route de Narbonne, F-31062 Toulouse, France} Manoel Manghi\affil{1}{} and Nicolas Destainville\affil{1}{}}

\contributor{Submitted to Proceedings of the National Academy of Sciences
of the United States of America}

\significancetext{Deciphering the physical mechanisms underlying the cell membrane organization at the nanoscale becomes conceivable with the rapid development of super-resolution fluorescence microscopy.  However the respective contributions of equilibrium and active membrane recycling processes driven by the cell in shaping the membrane remain controversial. Our theoretical approach demonstrates that switching off membrane recycling results in a significant two-fold decrease of typical nano-domain sizes, opening the route to new experimental strategies to ascertain the role of membrane recycling in this context.}

\maketitle

\begin{article}
\begin{abstract}
{Cell membranes are out of thermodynamic equilibrium notably because of membrane recycling, i.e. active exchange of material with the cytosol. 
We propose an analytically tractable model of biomembrane predicting the effects of recycling on the size of protein nanodomains. It includes a short-range attraction between proteins and a weaker long-range repulsion which ensures the existence of so-called cluster phases at equilibrium, where monomeric proteins coexist with finite-size domains. Our main finding is that when taking recycling into account, the typical cluster size increases logarithmically with the recycling rate. Using physically realistic model parameters, the predicted two-fold increase due to recycling in living cells is very likely experimentally measurable with the help of super-resolution microscopy.}
\end{abstract}


\dropcap{B}eing the interface between the cellular and extracellular media, the cell membrane has plenty of important biological functions such as signal transduction or transport of solutes~\cite{Alberts,Griffie16}. In the original cell membrane model by Singer and Nicolson (in 1972)~\cite{Singer1972}, the plasma membrane was visioned as an homogeneous mixture of lipids and proteins in which the proteins represent about 50~\% of the membrane mass. Since then, this basic model has known regular improvements showing an increasing organizational complexity. The organization in nano-domains had been suspected for long, but a variety of super-resolution light microscopy techniques and atomic-force microscopy (AFM) have recently improved further our understanding of membrane supramolecular organization. They gave definitive evidence of the generic existence of nanometer sized functional domains on the cell membrane~\cite{Griffie16,Scheuring2005,Betzig2006,Grage2011,Lang,Whited2015,Zuidscherwoude2015,RevueDestainLang}. 

From a biological perspective, many ideas on the functioning and roles of proteins clusters have been proposed~\cite{Griffie16}. First of all, clusters can be composed of different types of proteins and lipids that perform a certain biological task together and therefore need to be closely segregated in a tiny membrane region. A biologically important example for these specialized clusters is the G-protein clustered with receptors and effectors in signaling platforms~\cite{GPCRbook}. G-protein coupled receptors (GPCR) have been shown to be involved in a wide variety of bological processes by activating cellular signal transduction pathways~\cite{GPCRbook,GPCRs}. For example, the MOR-GPCR was examined by particle-tracking and was found to be locally confined in nano-domains, presumably because of inter-protein long-range forces~\cite{Destconfined}. E-cadherins constitute another example for which nano-patterning of the membrane has been shown to be involved in its biological function, clusters serving as adhesive foci in adherens junctions~\cite{Cavez2008,Truong2013}. Grouping identical receptors in a same cluster can also optimize and make more reliable their response to external stimuli~\cite{Grage2011,Whited2015,Mello2004,Fallahi2009,Gurry2009}. Clusters could also result in the inactivation or storage of proteins that would otherwise interact with their environment in a possibly unfavorable way. Forming domains could function as a mean of regulating the concentration of the proteins on the cell membrane, as proteins in the reserve pool would not be active~\cite{Lang}. 
\begin{figure}
\centering
    \includegraphics[width=8.8cm]{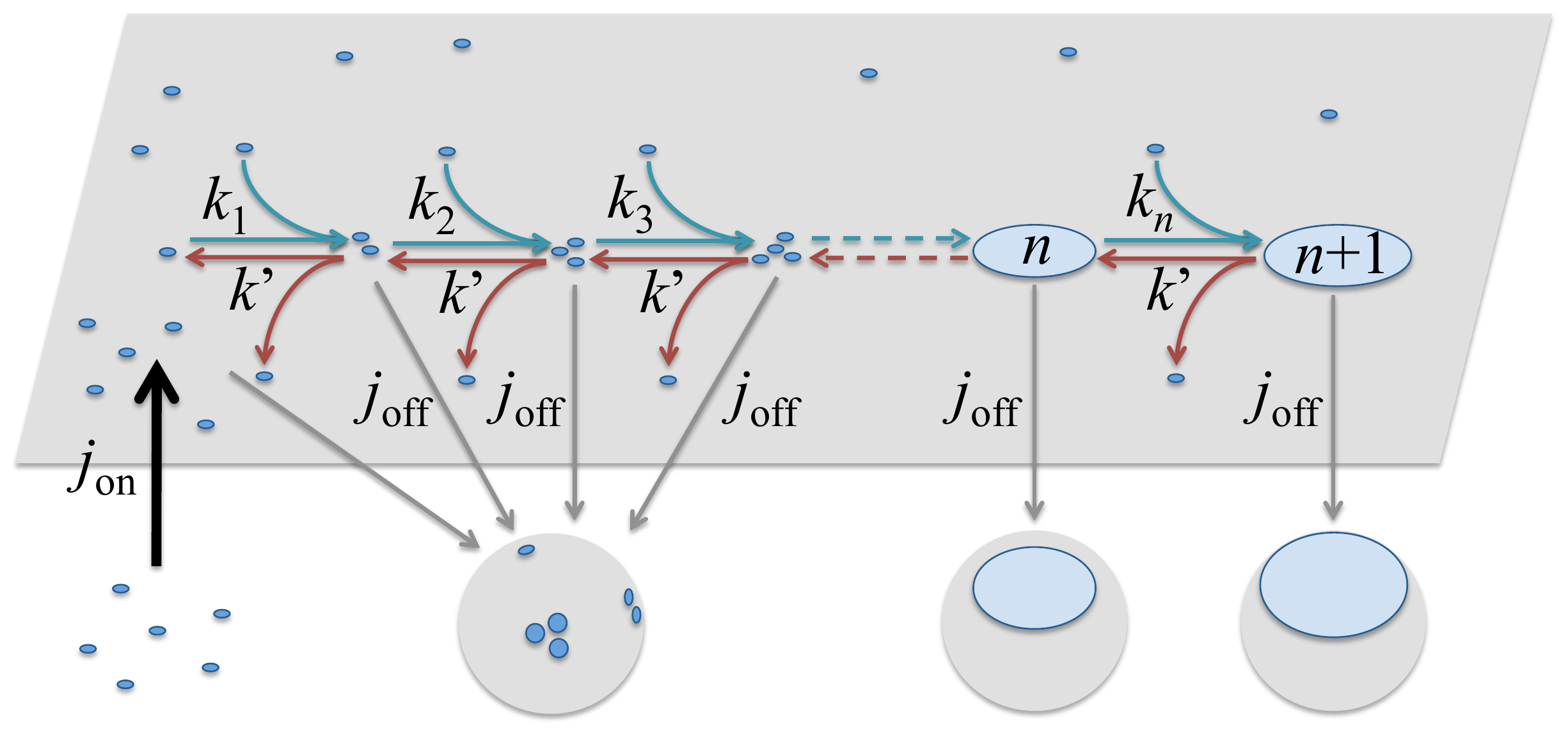}
\caption{Sketch of our out-of-equilibrium model of cluster dynamics: inside the membrane (in gray),  clusters can eject and capture monomers, with respective reaction rates $k'$ and $k_n$ for a cluster of size $n$. Monomers are injected in the system from the cytosol (black arrow on the left) with rate $j_{\rm on}$, whereas vesicles (gray spheres) are taking material from the membrane to the cell interior (gray arrows). Thus clusters are extracted from the membrane with a rate $j_{\rm off}$ independent of their size.
\label{turner}}
\end{figure} 

This article aims at developing a simplified biophysical model of the cell membrane describing the lateral distribution of embedded proteins and giving a realistic explanation of the formation of nano-clusters~\cite{RevueDestainLang,Lenne2009}.  In order for the proteins to be condensed in a cluster phase, a generic attractive interaction at short range is required, the origin of which has been discussed in many works (\cite{Grage2011,RevueDestainLang}~and references therein). Two types of theoretical approaches have been developed to account for the finite size of protein clusters~: (1)~an effective long range repulsive potential between the proteins that prevents the creation of one big macrophase \emph{at equilibrium}~\cite{Destain08,Gurry2009,Sieber2007,Destainville} and (2)~\emph{far-from-equilibrium} recycling of the cell membrane due to material exchange with the cytosol, that continually mixes the plasma membrane components and breaks large assemblies~\cite{Hancock2006,Jacobson2007,Turner,Fan2010B}. Here we combine these two approaches in a single analytically tractable theory. We demonstrate that taking out-of-equilibrium considerations into account does not change the qualitative properties of the model. However, switching recycling on increases the typical cluster sizes logarithmically with the recycling rate, which should be experimentally measurable.

\section{State of the art and model}

If a condensed phase exists below the critical temperature (or above the critical concentration), minimization of the interfacial free energy at equilibrium should lead to a large macro-phase in coexistence with a low density ``gas'' phase of monomers. This is in contradiction with the experimental observation of nano-domains in cells. Therefore, a mechanism is required which explains why condensation stops at a given domain size. Before presenting our own model in detail, we rapidly review useful anterior approaches tackling the finite size of nano-clusters (see Refs.~\cite{RevueDestainLang,Lenne2009,Fan2010B} for more details). 

At equilibrium, competition between the short-range attraction discussed above and a weaker but longer-range attraction can lead to nano-clustering by destabilizing too large protein assemblies. This is the so-called \emph{cluster phase mechanism}~\cite{Destain08}. The origin of the repulsion can be many-fold~\cite{RevueDestainLang}. An efficient mechanism comes from the up-down symmetry breaking that a vast majority of proteins are supposed to impose to the membrane, leading to a local spontaneous membrane curvature~\cite{RevueDestainLang}. The coupling between inclusion concentration and membrane curvature has recently received a striking experimental demonstration~\cite{Prevost2015}. When grouped in a cluster, asymmetric proteins collectively shape the membrane and lead to spherical buds, the elastic energy of which grows faster than the number of proteins~\cite{Weitz2013}. This is equivalent to the existence of a long-range repulsion of elastic origin which destabilizes too large clusters, because the energetic cost would overcome the gain in terms of line tension if macroscopic clusters grew~\cite{Sieber2007,Destain08,Gurry2009}. A typical finite cluster size emerges~\cite{Destainville,Weitz2013}, controlled by the attraction/repulsion competition.

Out-of-equilibrium active processes are also capable to explain the finite size of nano-domains~\cite{Hancock2006,Jacobson2007}. The coalescence of domains by diffusion, ultimately leading to a macro-phase at equilibrium, is in fact in competition with the traffic of membrane material to and from the membrane, for example by endocytosis and exocytosis of membrane patches. This membrane recycling breaks too large assemblies into smaller pieces. The growth of the condensed phase is thus stopped at a given typical size~\cite{Truong2013,Turner,Foret2005,Gomez2009,Fan2010A,Foret}. A simple dynamic scaling argument has for instance been proposed in Ref.~\cite{Foret2005}: when starting from a random configuration, one shows that the typical domain size grows with time as $L \approx (Dat)^{1/3}$ where $D$ and $a$ are the protein diffusion coefficient and typical diameter. Additionally, protein traffic sets a typical recycling time $\tau_r$, itself depending on off- and on-rates, from and to the membrane. Recycling prevents equilibration beyond the time scale $\tau_r$ and the coarsening should stop at a domain size $L \approx (Da\tau_r)^{1/3}$. 

Beyond this simplistic viewpoint, there are several ways to implement recycling~\cite{Turner,Fan2010B}. Proteins can arrive at the membrane as monomers through direct exchange with the cell cytosol. This is particularly true for peripheral proteins on the internal leaflet~\cite{Alberts}. Alternatively, they can be carried by vesicles, for example during exocytosis. There is no reason to anticipate that they arrive in clusters as large as those observed on the membrane and we assume for simplicity sake that they are principally carried as monomers, as in Refs.~\cite{Cavez2008,Foret}. Conversely, endocytosis and related processes remove material from the membrane. Without better insight, we assume that the removed membrane patches have the same composition as the average membrane one. The off-rate of any cluster type is thus assumed to be proportional to its surface fraction. Our recycling scheme is thus of ``monomer deposition/raft removal'' (MDRR) type as defined in Ref.~\cite{Turner}. These processes are assumed to be active, that is to say driven by energy consumption. The membrane is thus maintained out of equilibrium and we study its steady state.

 The goal of the present work is to quantify the respective contributions of equilibrium and far-from-equilibrium mechanisms in setting the typical cluster size. Using a formalism first proposed by Smoluchowski in a different but related context (see, e.g., Ref.~\cite{Family1986}), we will write a below master equation that embraces both equilibrium and out-of-equilibrium features. 

We adopt an approach where the plasma membrane is modeled as a two-dimensional fluid consisting of two types of particles. For the sake of simplicity, we indeed assume all proteins to be identical and the surrounding lipid phase to be homogeneous. We are interested in the distribution of the proteins in the lipid phase. Since proteins represent about one half of the membrane mass, the protein surface fraction $\phi$ is on the same order of magnitude, even though certainly slightly smaller because many proteins protrude out of the membrane plane. Accepted values lie around $\phi=0.2$~\cite{Whited2015,RevueDestainLang}. We shall explore the range $\phi = 0.01$ to 0.3 in this work and the SI. The proteins are clustered in $n$-particle multimers (Fig.~\ref{turner}) with $n=1$ for monomers. The cluster-size distribution is described by the surface fraction $c_n=\text{number of $n$-mers} \times s/S$
where $S$ is the system area. The area of a monomer is $s=\pi b^2\,$  with an approximate monomer diameter $2b=5$~nm~\cite{Alberts,Bionumbers}. 
If $N$ is the total number of proteins, their total surface fraction is 
\begin{align}
\phi \equiv N \, s/S  =\sum_{n=1}^\infty c_n\,n \, . 
\end{align}
The surface fractions of domains of all sizes ($n\ge1$) and of multimers ($n\ge2$) are respectively
$M=\sum c_n$ and 
$M^*= M - c_1$. The mean aggregation numbers of all sizes and of multimers are respectively
\begin{equation}
\bar{n}=\frac{\phi}{M}\; , \ \ \ \ \
n^*=\frac{\phi-c_1}{M^*}\; . 
\label{nstar}
\end{equation}

\section{Thermodynamical equilibrium in absence of recycling}

The cluster-size distribution $(c_n)$ at equilibrium is found by minimizing the free energy per particle~\cite{Safran}:
\begin{equation}
\frac{F}N = \bar f =\frac1{\phi} \; \sum\limits_{n=1}^{\infty}c_n\ln\left(c_n\right)-c_n+c_n\,F(n)\;.
\end{equation}
where the two first terms in the sum account for the mixing entropy. Here and in the following all energies are measured in units of the thermal energy $k_{\text{B}}\,T$ and the temperature $T$ is held fixed. 
The chemical potential conjugated to $N$ is
$\mu=\left({\partial F}/{\partial N}\right)_{S,T}$. The concentration ensues~\cite{Destainville,Safran}: $c_n=e^{\mu n-F(n)}$. The cluster-size distribution depends only on the chemical potential $\mu$ and the free energy of a domain $F(n)$. Following Ref.~\cite{Destainville,Foret}, we  use the following form
\begin{equation}
F(n)  =  -f_0 (n-1) + \gamma \sqrt{n-1} + \chi (n-1)^\alpha.
\label{Fden}
\end{equation}  
As explained in the SI (Eq.~\eqref{Fden:SI}), it combines a classical liquid droplet approximation in two dimensions (the first two terms) and a repulsive contribution leading to the finite size of clusters at equilibrium (the last term)~\cite{RevueDestainLang}. The value of $\alpha$ has been thoroughly explored~\cite{Weitz2013} and typically lies between 1.5 and 2, the exact value depending on the membrane tension. Below we first choose $\alpha=2$, the low tension case where analytical calculations can be performed entirely. The case $\alpha=3/2$ is tackled numerically in the SI (Fig~\ref{alpha=2}), and we arrive to similar qualitative conclusions in this case.

The order of magnitudes of the parameters have already been discussed in connection with biophysical data~\cite{Destain08,Destainville,Weitz2013}: The value of $f_0=-(f_{\rm att}+f_{\rm rep})>0$, in the 10 to 30~$k_BT$ range, is in fact of little interest because it is $f_0+\mu$ which eventually determines the value of $\phi$; $\gamma$ has the same order of magnitude as $f_0$. If we choose $\alpha=2$, $\chi$ appears to be of order 0.1 to get equilibrium clusters containing few dozens of particles as observed experimentally. We thus choose $f_0=20$, $\gamma=20$ and $\chi=0.1$ (in units of $k_BT$) as a reference parameter set. Alternative parameter sets are studied in the SI, with similar qualitative conclusions. 

\begin{figure}[h!]
\centering
\includegraphics[width=7.5cm]{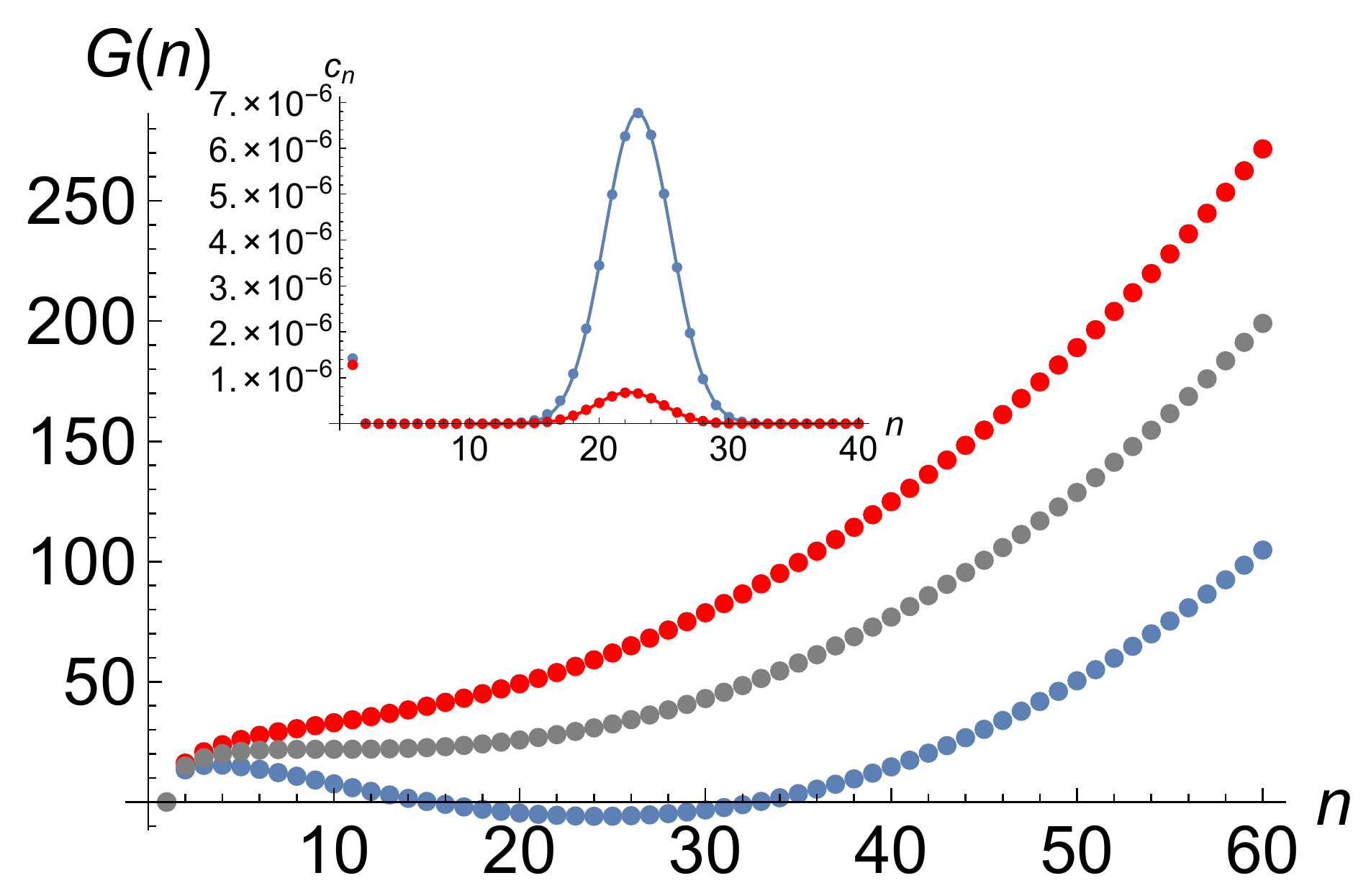}
    \caption{$G(n)$ (in units of $k_{\rm B}T$) for the reference parameter set $\alpha=2$, $f_0=20$, $\gamma=20$ and $\chi=0.1$. Top curve (in red): Before the phase transition, here for $\phi = 10^{-7}$, the energy for monomers is minimal. Middle curve (in gray): For $\phi=\phi_c \simeq 3.48 \times 10^{-7}$, $\tilde{G}(n)$ has an inflection point near $n=10$. Bottom curve (in blue): Above the transition, here for $\phi=0.1$, there are two local minima of comparable energy, one for monomers and one for higher aggregation numbers. Inset: Surface fractions $c_n$ of clusters of size $n$ (or $n$-mers) at equilibrium, for two different surface fractions $\phi=10^{-4}$ (red dots) and $10^{-3}$ (blue dots). Monomer concentrations ($n=1$) are nearly superimposed. Note that the multimer peak position increases very slowly with $\phi$, as well as the monomer surface fraction $c_1$. Gaussian fits of the multimer peak are superimposed (continuous lines). 
    \label{transrep}}
\end{figure}

In order to compute the $c_n$, we switch now to the grand-canonical ensemble. It can be shown~\cite{Safran} that the adequate thermodynamical potential is now $\tilde G(n)=F(n)-\mu n$. Using $c_n=e^{\mu n-F(n)}$ and the fact that $F(1)=0$, we get
\begin{equation}
c_n=e^{-\tilde{G}(n)} =c_1\,e^{-G(n)} 
\label{exact:eq}
\end{equation}
with $\mu=\ln(c_1)$, after introducing $G(n)=\tilde{G}(n)+\mu$, i.e.
\begin{align}
G(n) = -(f_0+\mu) (n-1) + \gamma \sqrt{n-1} + \chi (n-1)^\alpha.
\end{align}
The value of $\mu$ (or equivalently of $c_1$) is fixed by the value of $\phi$ through the constraint $\phi=\sum c_n n$~\cite{Destainville}. Fig.~\ref{transrep} shows $G(n)$ for different values of $\phi$. For a very low $\phi$ (or very low $\mu$), the energy for forming a monomer is minimal. At the critical concentration $\phi_c$,  $\tilde{G}(n)$ has an inflection point indicating the existence of a local minimum for higher values of $\phi>\phi_c$ where clusters nucleate.

Above $\phi_c$ the cluster-size distribution is bimodal. Monomers, and very rare dimers or trimers, co-exist with a condensed phase of larger clusters of average size $n^*$. From the inset of Fig.~\ref{transrep}, it can be seen that the multimer peak is well fitted by a Gaussian centered on $n^*$ (see Eq.~\eqref{nstar})
\begin{equation}
c_n = A_0 \exp \left[  -\frac{(n-n^*)^2}{2\sigma_0^2} \right]
\label{Gaussian}
\end{equation}
for $n$ close to $n^*$.
Note that $n^*$ is nearly insensitive to $\phi$~\cite{Destainville} because the second minimum of $\tilde{G}(n)$ itself is slowly shifted to the right when $\mu$ grows.

\section{Steady state in presence of recycling}

The previous model gives one explanation of the occurrence of finite-size clusters appearing on the cell membrane. Still the cell membrane  constantly exchanges material with the cytosol, which we now take into account. The following master equation describes the time-evolution of the domain size distribution~\cite{Family1986,Turner,Truong2013}:
\begin{align} \label{mastergeneral}
\frac{{\rm d}c_n}{{\rm d}t}&=\mathcal{J}(n)+\sum\limits_{m=1}^{\infty} \left( k_{n,m}\,c_{n+m}-k'_{n,m}\,c_n c_m \right) \\
\nonumber&+\frac12 \sum\limits_{m=1}^{n-1} \left(k'_{m,n-m}\,c_{n-m} c_m-k_{m,n-m} c_n \right) \; ,
\end{align}
where $\mathcal{J}(n)$ is the $n$-cluster net flux from the cytosol to the membrane ($\mathcal{J}=0$ at equilibrium). Equilibrium concentrations will be denoted by $c_n^{(0)}$  from now on. For monomers, the second sum on the r.h.s vanishes
because monomers cannot result from the fusion of smaller clusters. This mean-field approach is presumably valid in dimension 2~\cite{Family1986}.

The coefficients $k_{n,m}$ and $k'_{n,m}$ control the rates  of cluster fragmentation, in which one domain of $n+m$ particles breaks into two smaller ones of size $n$ and $m$, and domain fusion, in which two domains containing $n$ and $m$ monomers fuse to form a single domain of size $n+m$ \cite{Turner}. At equilibrium, the system makes in average equally often the transition from one state to the other, which translates into he detailed balance condition $c_{n+m}\,k_{n,m}=k'_{n,m}\,c_{n}\,c_{m}$~\cite{vanKampen}.

We assume from now on that only the fragmentation and fusion terms involving monomers  are taken into account, as in Ref.~\cite{Foret}. Indeed, the energy barrier to be overcome by a dimer or a small multimer to escape a cluster of size $n \gg 1$ is significantly larger than the one for a monomer because the binding energy between a small $m$-mer and a large $n$-mer is roughly proportionnal to $m$. Owing to Kramers' theory,  $m$-mers with $m >1$ can hardly escape from a cluster. Conversely, such multimers are very rare in the ``gas'' phase and their fusion with clusters is therefore exceptional. This argument is corroborated by the absence of observation in kinetic Monte Carlo simulations of fragmentation and fusion events involving $m$-mers with $m >1$~\cite{Destain08}. The master equation simplifies to
\begin{align}
\frac{{\rm d}c_n}{{\rm d}t}= \mathcal{J}(n) +k_nc_{n+1}-k'c_nc_1 
+k'c_{n-1}c_1-k_{n-1}c_n
\label{master}
\end{align}
for $n\geq 2$ where $k_n \equiv k_{n,1}=k'\,{c_n^{(0)}\,c_1^{(0)}}/{c_{n+1}^{(0)}}$
because the kinetic constants are assumed to be equal to their equilibrium counterparts~\cite{vanKampen}.
For monomers, 
\begin{align}
\frac{{\rm d}c_1}{{\rm d}t} = \mathcal{J}(1) - k'\, M c_1 
+\sum_{m=1}^{\infty} k_m\,c_{m+1}
\label{master:c1}
\end{align}
where we recall that $M=\sum_{n \geq 1} c_n$.

We have chosen to consider $k'$ as independent of $n$ because it measures the capture rate of a monomeric particle by a $n$-cluster through a diffusive process. Either the capture process is reaction-limited or it is diffusion-limited. In the former case, the capture reaction is limited by a repulsive energy barrier between the cluster and the monomer that the monomer has to overcome. Since a vast majority of clusters have a size $n$ close to $n^*$, the energy barrier can be considered as essentially independent of $n$. In the latter case, solving the 2D backward Smoluchowski equation shows that the capture time is  $\tau \simeq b^2/(2D)  \ln(b/a)$ where $a$ is the cluster diameter and $b>a$ the typical distance between clusters; $D$ is the diffusion coefficient of a free monomer in the bilayer~\cite{Condamin2005}. For the same reason as above, $b$ and $a$ can be considered as essentially independent of $n$. In both cases, $k'$ weakly depends on $n$ and will be considered as constant. If the capture is diffusion-limited, taking $D\sim 10^{-12}$~m$^2$s$^{-1}$~\cite{Bionumbers}, $a\simeq50$~nm~(\cite{RevueDestainLang} and references therein) and $b\simeq200$~nm (so that $\phi \approx a^2/b^2 \approx 0.1$), we get $\tau \sim 0.01$~s. Now owing to the master equation~\eqref{master:c1}, $\tau$ is related to $k'$ through $\tau^{-1} = k' M_0$~[\footnote{The capture time $\tau$ is also the typical time of the fusion reactions $(n) + (1) \rightarrow (n+1)$ in Eq.~\eqref{master:c1}, at equilibrium, if fission and recycling terms were absent, that is to say of the differential equation ${\rm d}c_1 / {\rm d}t = - k'\, M_0 c_1$. Hence $\tau^{-1} = k' M_0$.}]  
where $M_0 = \sum_{n \geq 1} c_n^{(0)} \approx \sqrt{2\pi} \sigma c_{n^*}^{(0)} \simeq 10^{-3}$ is the cluster concentration (estimated by the integration of the Gaussian). We get $k'\sim 10^5$~s$^{-1}$ and we shall use this value in the following. Ref.~\cite{Turner} proposes the same estimate of $k'$ by using a related argument. 

As justified above, we consider in this work the ``monomer deposition/raft removal'' (MDRR) scheme of Ref.~\cite{Turner} (Fig.~\ref{turner}). Endocytosis takes clusters from the membrane with a rate $j_{\rm off}$ independent of their size and proteins are assumed to be transported to the membrane in a monomeric state with a rate $j_{\text{on}}$. Thus the recycling term reads $\mathcal{J}(n)=j_{\text{on}}\,\delta_{n,1}-j_{\text{off}}\,c_n$~\cite{Turner}. At steady state, the total recycling term is ${\rm d}\phi / {\rm d} t = 0 = \sum_{n=1}^{\infty}\mathcal{J}(n)\,n$ and
\begin{align}
\phi=j_{\text{on}}/j_{\text{off}}\, .
\end{align}
The surface fraction $\phi$ is now fixed by this constraint and not by the equilibrium chemical potential $\mu$ anymore. 

At steady state, all concentrations satisfy ${\rm d}c_n/{\rm d}t=0$. The master equation~\eqref{master} thus provides an order-2 recurrence relation relating $c_{n+1}$, $c_n,$ and $c_{n-1}$. Solving this recurrence requires the knowledge of $c_1$ and $c_2$, and in addition the master equation for $n=1$ involves all the $c_n$. This is not a practical way to tackle the problem. Taking a maximum aggregation number $n_{\rm max}$ much larger than the typical cluster size, an order-2 {\em descending} recurrence can be built up instead. 

In the $j_{\rm off} \rightarrow 0$ limit, the detailed balance condition is close to be respected. It ensues that $k_n \simeq k'\,c_n\,c_1/c_{n+1}$.
Comparing this relation to its equilibrium counterpart, we get $c_{n+1}/c_{n+1}^{(0)} \simeq (c_1/c_1^{(0)})(c_n/c_n^{(0)})$. We are thus led to write 
\begin{equation}
c_n=c_n^{(0)} \left[ \frac{c_1}{c_1^{(0)}} \right]^n g(n),
\label{cnvscn0}
\end{equation}
where $g(n)$ is some factor to be determined, satisfying $g(1) \equiv 1$ by definition and $g(n)\rightarrow 1$ when $j_{\rm off} \rightarrow 0$ for any $n$. Injecting this identity in Eq.~\eqref{master}, the descending recurrence relation
\begin{equation}
g(n-1) = \frac{\left[ \hat j_{\rm off} + c_1 + c_1^{(0)} \Gamma(n)  \right] g(n) - c_1 g(n+1)}{c_1^{(0)} \Gamma(n)}
\label{desc:rec}
\end{equation}
ensues at steady-sate for $n\geq 2$, after introducing the notations $\hat j_{\rm off} = j_{\rm off} /k'$ and 
\begin{equation}
\Gamma(n) \equiv \frac{c_{n-1}^{(0)}}{c_n^{(0)}} = e^{G(n)-G(n-1)} \simeq   e^{G' (n)}.
 \label{def:Gamma}
 \end{equation}

\section{Numerical results}

\begin{figure*}[t!]
\centering
\includegraphics[width=\textwidth]{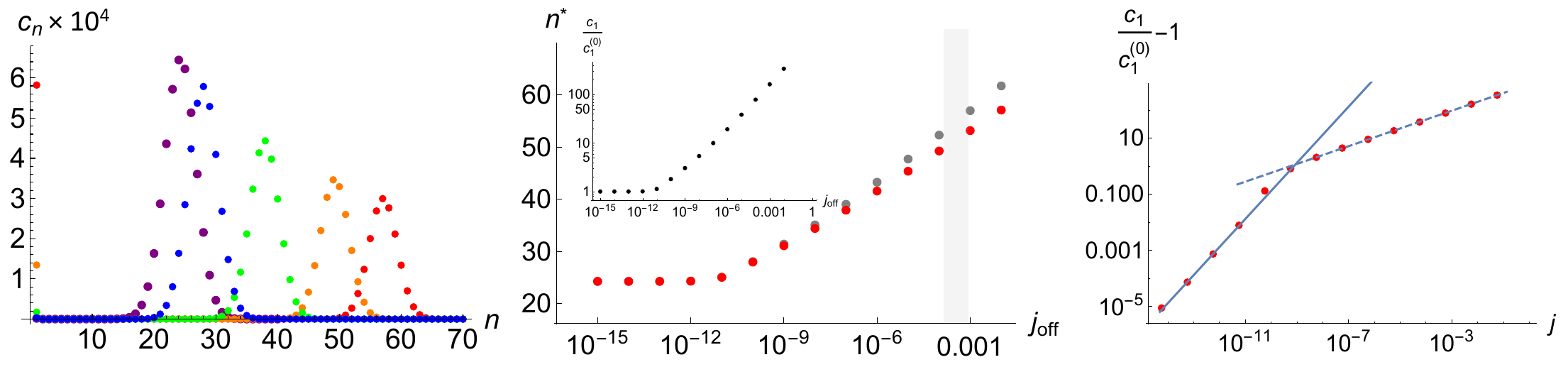}
\caption{The parameters are the reference ones ($\alpha=2$, $f_0=20$, $\gamma=20$ and $\chi=0.1$) and $\phi=0.1$. Left: Concentrations (or surface fractions) $c_n$ of $n$-clusters from the thermodynamical equilibrium state to high recycling rates. The recycling rates (in s$^{-1}$) are given by $j_{\text{off}}=0$ (equilibrium, in purple),  $10^{-10}$ (blue), $10^{-7}$ (green), $10^{-4}$ (orange) and $10^{-2}$ (red). The large-$n$ cutoff was $n_{\rm max}=150$ in theses numerical calculations.
Middle: Exact values of $n^*$ as calculated through the descending recurrence (red dots) as well as the approximate value inferred from the values of $c_1$ and  Eq.~\eqref{nstarorder1} (gray dots, superimposed with the red ones for low values of $j_{\rm off}$). The gray rectangle shows the biologically relevant values of $j_{\rm off}$. Inset: $c_1/c_1^{(0)}$ in function of $j_{\rm off}$ (in s$^{-1}$). Log-linear (main panel) and log-log (inset) coordinates.
Right: Comparison of the monomer concentration $c_1$ vs the reduced recycling rate $j$ found numerically (dots; same data as in the inset of the middle figure) and analytically through the small $j$ expansion (continuous line). From the notations defined in the text, we expect $c_1/c_1^{(0)}-1= [c_1^{(1)}/c_1^{(0)}] \, j + \mathcal{O} (j^2)$. This scaling ceases to be valid above a limiting value $j\sim 10^{-9}$.  The dashed line is then a power-law fit of the six last dots $c_1 \propto j^\nu$ with $\nu \simeq 0.32$ (Eq.~\eqref{scaling:nu}). Log-log coordinates. 
\label{c1:joff}
}
\end{figure*}

The numerical method used to solve Eq.~\eqref{desc:rec} is given in the SI. The so-obtained distributions $(c_n)$ are displayed in Fig.~\ref{c1:joff} (left), and show that the multimer distributions remain Gaussian. The recycling rate in a real cell plasma membrane is estimated as follows. Experiments indicate that the typical amount of plasma membrane endocytosed per hour is equivalent to the total surface of the cell plasma membrane of a fibroblast~\cite{Steinman1983}. The turnover time of plasma membranes is thus typically of 1~hour. Now $1/j_{\rm off}$ is precisely the typical time during which all clusters of any size have been recycled. Thus  $1/j_{\rm off}\sim 1$~hour and $j_{\rm off}\sim 10^{-4}$ to $10^{-3}$~s$^{-1}$~\cite{Foret}. The numerical results then show that the average cluster size $n^*$ increases with $j_{\rm off}$, by a factor $\simeq 2$ between equilibrium and these biological recycling rates.

The inset of Fig.~\ref{c1:joff} (middle) shows how $c_1/c_1^{(0)}$ also grows with recycling. Qualitatively, when $j_{\rm on}$ (and $j_{\rm off}$) grow, more monomers are added to the system, which tends to increase the value of $c_1$: $c_1>c_1^{(0)}$. The  reactions $(n) + (1) \leftrightarrows (n+1)$ are displaced to their right, because the flux from the left to the right is proportional to $c_1$. This tends to increase the relative fraction of larger multimers as compared to their equilibrium counterparts. The Gaussian peak is shifted to the larger values of $n$. Owing to Eqs.~\eqref{Gaussian} and \eqref{cnvscn0}, approximating $g(n)$ by a constant near $n_0^*$, we get $c_n \simeq \text{Const.}  \exp [-(n-n^*_0)^2/2\sigma_0^2 + n \ln (c_1/c_1^{(0)})] $ where $n^*_0$ is the equilibrium mean multimer size. By completing the square in the exponential, it follows that the new maximum of $c_n$ is at
\begin{equation}
n^* = n^*_0 + \sigma_0^2 \ln (c_1/c_1^{(0)}).
\label{nstarorder1}
\end{equation}
Numerical values are consistent with this calculation while $j_{\rm off}<10^{-8}$~s$^{-1}$ (see Fig.~\ref{c1:joff}, middle). When $j_{\rm off}$ gets larger, this approximation becomes less reliable.

\section{Analytical solution}

In the following we propose analytic identities relating the 4 parameters that characterize the distributions $(c_n$), namely the monomer fraction $c_1$, the average multimer size $n^*$, the width $\sigma$ of the Gaussian peak and its amplitude $c_{n^*} \equiv A$. We assume that the distribution can be written as a Gaussian plus a delta peak at $n=1$ accounting for monomers
\begin{align}
c_n \simeq c_1\,\delta_{n,1} + A\exp \left[ \frac{-\left(n-n^*\right)^2}{2\,\sigma^2} \right].
\label{cn}
\end{align}
As seen in Fig.~\ref{c1:joff} (left), the transition from the equilibrium state to low recycling rates at fixed $\phi$  is performed by smoothly shifting the Gaussian peak towards higher aggregation numbers while in the same time the number of monomers is increasing because monomers are injected in the system through recycling. The width $\sigma$ is very slowly depending on $j_\text{off}$. 

We focus on the concentrations $c_n$ at aggregation numbers $n$ close to $n^*$. Inserting Eq.~\eqref{cn} in the master equation~\eqref{master} at steady state gives for any $n$ close to $n^*$:
\begin{eqnarray}
\hat j_{\text{off}}&=& e^{\gamma\,\left(\sqrt{n}-\sqrt{n-1}\right)-f_0+ \chi\,\left(n^\alpha-\left(n-1\right)^\alpha\right)-\frac{1+2\,\left(n-n^*\right)}{2\,\sigma^2}}   \nonumber \\
&& -c_1+ c_1\,e^{-\frac{1-2\,\left(n-n^*\right)}{2\,\sigma^2}} \nonumber \\
&& - e^{\gamma\,\left(\sqrt{n-1}-\sqrt{n-2}\right)-f_0+\chi\,\left(\left(n-1\right)^\alpha-\left(n-2\right)^\alpha\right)}  \, .
\label{master2}
\end{eqnarray}
The expressions are linearized in $n-n^*$, assuming that 
\begin{equation}
\frac{n-n^*}{n^*}\ll1 \ \mbox{ and } \ \frac{n-n^*}{\sigma^2}\ll1\, .
\label{conds}
\end{equation}
These approximations can be justified by the observation that $n^* \gg 1$ in general and especially for high recycling rates and that the standard deviation was found in the range $\sigma\simeq 2.5$ to 6 for the various parameter sets explored in this study.

By equating the so-obtained expansion to 0, both order-0 and order-1 coefficients must vanish simultaneously. As detailed in the SI (Eqs.~\eqref{taylor:exp1}~and below), it follows from the order-1 coefficient that
\begin{align}
c_1 \simeq \exp \left[ F'(n^*) \right] = c_1^{(0)} \Gamma(n^*)\, .
\label{c1BIS} 
\end{align}
The second equality ensues from  $F'(n)=G'(n) + \mu$ and $\exp(\mu)=c_1^{(0)}$. It can also be checked that Eq.~\eqref{nstarorder1} derives simply from the expansion of this relation at order one in the small variable $c_1/c_1^{(0)}-1$. In the same way we find in the SI an expression for $\sigma$ by equating the order-0 term to zero: 
\begin{equation}
\sigma \simeq \left[ 2\chi - \frac{\gamma}4  \frac1{\left( n^* \right)^{3/2}}  \right]^{-1/2}.
\label{sigm2} 
\end{equation}
The next step is to find an expression for the amplitude $A$. We use the condition $\phi=\sum c_n n$ and Eq.~\eqref{cn}:
$\phi=c_1+A\,\sum\limits_{n=2}^{\infty}n \, e^{\frac{-\left(n-n^*\right)^2}{2\,\sigma^2}} \simeq c_1+\sqrt{2\,\pi}\,A\,n^*\,\sigma$.
It follows that 
\begin{align}
A &=\frac{\phi-c_1}{\sqrt{2\,\pi}\,\sigma\,n^*}.
\label{eqA}
\end{align}
At low or moderate recycling, $c_1 \ll \phi$ can be omitted in the numerator. In Figs.~\ref{figc1}, \ref{figsigm} and \ref{A}, we compare the analytical values of the monomer concentration $c_1$, the standard deviation $\sigma$ and the amplitude $A$ with the values found by fitting the numerical distributions of Fig.~\ref{c1:joff} (left) with a Gaussian. The very good agreement shows the validity of our approximations. Our next step will consist in deriving an analytical expression for these different quantities in function of $\hat j_\text{off}$.

\emph{Small $\hat j_{\rm off}$ limit}~-- At equilibrium ($j_{\rm off}=0$) Eq.~\eqref{cnvscn0} implies that $g_{\rm eq}(n) = 1$ for all $n$. Our small parameter will be $j \equiv j_{\rm off}/(k' c_1^{(0)})$. We naturally write $g(n)=1 - w_n  j + \mathcal{O}( j^2)$
where the sequence $(w_n)$ is to be characterized. Expanding Eq.~\eqref{desc:rec} at order 1 in $j$ leads to the recurrence relation
\begin{equation}
w_{n-1} = w_n\left(1+\frac1{\Gamma(n)}\right)  - \frac{w_{n+1}}{\Gamma(n)} + \frac1{\Gamma(n)}  \, .
\label{wn:rec}
\end{equation}

If one also expands $c_1=c_1^{(0)} + c_1^{(1)} \, j + \mathcal{O}(j^2)$, we show in the SI that if $w_\infty$ is the limit of $(w_n)$, then
\begin{equation}
c_1^{(1)} \simeq \frac{c_1^{(0)}}{n_0^*} \, w_{\infty} \, .
\end{equation}
The value of $w_{\infty}$ only depends on $G(n)$. With our parameters at $\phi=0.1$, $w_{\infty} \simeq 3.09  \times 10^{10}$, $c_1^{(0)} = 1.72 \times 10^{-6}$ and $n_0^*\simeq 24.2$. The obtained value $c_1^{(1)} \simeq 2.19 \times 10^3$ coincides within 1~\% with the value extracted from numerical calculations at very small recycling rates, $j_{\rm off}=10^{-15}$ to $10^{-12}$~s$^{-1}$, as illustrated in Fig.~\ref{c1:joff} (right). This again validates our approximations. Once $c_1$ is known, the other quantities of interest, $n^*$, $A$ and $\sigma$ can be expressed in function of $j$ by the three  relations in Eqs.~\eqref{c1BIS} to~\eqref{eqA}, and the problem is solved.

\emph{Intermediate values of $\hat j_{\rm off}$}~-- We start from Eq.~\eqref{cnvscn0s} in the SI where we make the substitution $c_1/c_1^{(0)} = \Gamma(n^*)$. We get $A=c_{n^*}^{(0)} \left[ \Gamma(n^*) \right]^{n^*} g(n^*)$. Next, we use $c_{n^*}^{(0)}=c_1^{(0)}e^{-G(n^*)}$. In the SI we also study analytically the sequence $g(n)$ and we prove (see Eq.~\eqref{last:rel}) that
$g(n^*)\simeq g(\tilde n) \simeq \kappa \left[ \Gamma(n^*) \right]^{-\tilde n} e^{G(\tilde n)}$ where 
\begin{equation}
\tilde n \simeq \tilde n_1 + (1-\chi) n^* + (\ln j)/2. 
\label{ntilde:eq}
\end{equation}
Here $\kappa>0$ and $\tilde n_1<0$ are constants independent of $n^*$ and $j$.  Furthermore, Eqs.~{\bf [\,20\,]} 
and \eqref{eqA} provide the approximation $A \simeq (\phi \sqrt{\chi})/(\sqrt{\pi} \, n^*)$ at large values of $n^*$. Altogether, these relations lead to 
\begin{equation}
K \simeq {n^*} \left[ \Gamma(n^*) \right]^{n^*-\tilde n} e^{G(\tilde n)-G(n^*)}
\label{K:eq}
\end{equation}
where $K= (\phi \sqrt{\chi})/(\sqrt{\pi} \kappa \, c_{1}^{(0)})$ is a constant. $K \simeq 5 \times 10^6$ with our reference set of parameters and $\phi=0.1$.  Eqs.~\eqref{ntilde:eq} and \eqref{K:eq} relate $\tilde n$ and $n^*$, which are both unknown. Solving this two-equation system gives both $\tilde n$ and $n^*$. This calculation is done in the SI, and we get
\begin{equation}
n^* \simeq {\rm Const.} +  \frac{\nu}{2 \chi}  \ln j
\label{log:scaling}
\end{equation}
with 
\begin{equation}
\nu = \left[ 1 + \frac{3\gamma}{32\chi} \sqrt{\ln \frac{K}{n^*_0}} \frac1{(n_0^*)^{1/4}\left[ \chi (n_0^*)^{3/2} -\frac{\gamma}8 \right]^{3/2}} \right]^{-1} .
\label{nu:main}
\end{equation}
Owing to Eq.~\eqref{c1BIS}, $c_1 \propto \Gamma(n^*)$ with $\Gamma(n^*) \simeq e^{-(f_0+\mu)} e^{2 \chi  n^*}$. One eventually gets the expected scaling law 
\begin{equation}
c_1 \propto e^{2 \chi  n^*} \propto j^\nu.
\label{scaling:nu}
\end{equation}
The searched exponent $\nu$ (see Fig.~\ref{c1:joff}, right) is given by Eq.~\eqref{nu:main}. With the reference parameter set, $\nu_{\rm analytic} \simeq 0.49$, whereas the numerical solution gave $\nu_{\rm num.} \simeq 0.32$. This overestimate by 50~\% is due to the various approximations we used. In the different parameter sets tested in the SI, $\nu_{\rm num.}$ remains close to 0.3. Note that the exponent $\nu$ would be difficult to measure experimentally because it would require to have access to the monomer density in the dilute phase, whereas only clusters are visible with super-resolution microscopy. In this respect, our main finding is that the typical cluster size $n^*$ grows logarithmically with $j$ (i.e. with $j_{\rm off}$) in this intermediate regime as well (Eq.~\eqref{log:scaling}).

The crossover between the small and intermediate values of $\hat j_{\rm off}$ (Fig.~\ref{c1:joff}, right) ensues from Eq.~\eqref{ntilde:eq}, where $\tilde n$ must be positive. When decreasing the value of $\hat j_{\rm off}$, both $n^*$ and $\ln j$ decrease. Thus Eq.~\eqref{ntilde:eq} ceases to be valid below a limiting value of $\hat j_{\rm off}$ and one crosses over the the small $\hat j_{\rm off}$ limit.

\section{Conclusion and discussion}

We have characterized several dynamical regimes depending on the strength of the recycling rates $j_{\rm off}$ or $j_{\rm on}$. (i)~In the low recycling regime, where $c_1$ is close to its equilibrium value $c_1^{(0)}$, i.e. where $(c_1-c_1^{(0)})/c_1^{(0)} \lesssim 1$, this quantity grows linearly with $j_{\rm off}$ and the proportionality coefficient can be calculated perturbatively. (ii)~In the experimentally relevant intermediate regime $c_1 \ll c_1^{(0)} \ll \phi$, we observed a power-law $c_1 \propto j_{\rm off}^\nu$ where the exponent $\nu \approx 0.3$ could be correctly estimated by an analytical argument. The biologically relevant range of recycling rates belongs to this regime. In these two regimes, the typical multimer size $n^*$ grows logarithmically with the recycling. (iii)~When increasing further the recycling rate, $c_1$ saturates to $\phi$ and proteins are essentially present in their monomeric form, thus destroying the cluster phase. 

The starting point of the present study was the existence of two types of physically relevant arguments accounting for the finite size of clusters in the condensed phase. Some works argue that even though a membrane is maintained out-of-equilibrium by active processes, equilibrium statistical mechanics is sufficient to account for the finite size of membrane nano-domains~\cite{Sieber2007,Destain08,Destainville,Gurry2009}. For example, the observation of nano-domains of rhodopsin (a GPCR) in cadavers' retina cells~\cite{Whited2015}, very similar to those observed in fixed cells or membrane sheets, suggests that active processes are not essential. Alternatively, some other works propose that out-of-equilibrium mechanisms are necessarily at play, driven by active membrane recycling~\cite{Turner,Foret2005,Cavez2008,Gomez2009,Fan2010A,Fan2010B,Foret}. We explored whether both approaches could be discriminated from an experimental perspective. Our conclusion is that the sole observation of cluster size distributions, as extracted from super-resolution microscopy experiments, cannot bring a definitive answer if one does not compare directly live cells and systems where recycling has been switched off. Within the realistic parameter sets that we have explored in this work, it appears that the typical cluster size $n^*$ is about twice smaller when switching recycling off, from a realistic value $j_{\rm off}\sim 10^{-3}$~s$^{-1}$ to 0. Thus our prediction is that stopping active recycling processes in a living cell in some way would result in twice smaller, and consequently twice more numerous clusters at fixed protein concentration. This should be measurable by super-resolution fluorescence microscopy or AFM. 

Our study relies on some assumptions, among which the ``monomer deposition/raft removal'' recycling scheme. Alternative recycling schemes have been discussed elsewhere~\cite{Turner,Fan2010B,Foret2005,Gomez2009,Foret}. However, we have argued that the chosen recycling scheme is realistic. Note also that the proximity of a phase transition can be relevant in the present context~\cite{Gueguen2014}. We have chosen here to focus on the simplest case of a system that is far from this transition (and below the transition temperature) in order not to obscure the interplay between equilibrium and out-of-equilibrium processes of interest. Combining equilibrium and out-of equilibrium considerations 
close to a phase transition potentially leads to an even richer phenomenology.



%


\newpage

\appendix[Supplementary Information]
\setcounter{figure}{0}
\makeatletter 
\renewcommand{\thefigure}{S\@arabic\c@figure}
\makeatother

\section{Cluster free energy}

\subsection{Classic liquid droplet approximation}

Below the critical temperature $T_c$, the line tension along the contact line between two phases (protein clusters and surrounding lipids) results from the fact that it is energetically unfavorable for the different types of particles to be directly next to each other (Ref.~\cite{Safran} of the main text). The line tension $ \hat \gamma >0$ acts on the cluster boundaries and results on a free-energetic cost $F^{\rm line}_n =\hat \gamma\,\left(2\,\pi\,r_n\right) \simeq \gamma\,\sqrt{n}$, with $r_n$ the radius of a roundish cluster of aggregation number $n$. The value of $\gamma$ is typically in the range of few tens of $k_{\text{B}}\,T$. For $\gamma=20$~$k_BT$ this yields $\hat \gamma\simeq 1.3\,k_BT$/nm~$\simeq 5$~pN for a monomer diametre $2b=5$~nm (see Refs.~\cite{Turner,Weitz2013} of the main text). Note that this $k_BT$/nm~$\sim$~pN order of magnitude is also comparable to the measured line tensions along the phase boundaries between different lipid phases in lipidic membranes~\cite{Riviere1995,Sriram2012}, in the alternative case where the phases of solute and solvent correspond in fact to two lipid phases. 

The roundish character of a droplet results from the minimization of its interfacial energy. Its energy is composed of the above boundary term plus a bulk contribution~\cite{Destainville}:
\begin{align}
F_{\rm att}(n)=f_{\rm att}\,\left(n-1\right)+\gamma\,\left(n-1\right)^{1/2}\; .
\end{align}
The free energy per particle $f_{\rm att}<0$ accounts for the short-range attractive potential between adjacent particles. We have chosen to express $F_{\rm att}(n)$ in function of $n-1$ (rather than $n$) so that $F_{\rm att}(1)=0$, as there is no binding energy in a monomeric cluster. When starting from a dilute gas phase and increasing the monomer concentration, such a free-energy leads to a macro-phase separation above a critical concentration. Then the essentially monomeric, low-density phase coexists with a large condensed phase, a giant cluster minimizing the surface energy term.

\subsection{Repulsive potential}

In addition to the energy of a liquid droplet, other energies can be involved. The existence of an effective long-range repulsion between the solute particles has been evoked in the main text. Contrary to the macro-phase separation above, this repulsion results in the existence of a stable phase of intermediate sized clusters at equilibrium. The long-range repulsive energy can be approximated by the combination of a linear term and a power law:
\begin{align}
F_{\rm rep}(n)=f_{\rm rep}\left(n-1\right)+\chi\left(n-1\right)^\alpha.
\end{align}
with an effective exponent $\alpha>1$, and where $\chi$ gives the strength of the repulsive potential. Combining the attractive and repulsive contributions in $F(n) = F_{\rm att}(n)+F_{\rm rep}(n)$ leads to 
\begin{eqnarray}
F(n) & = & -f_0 (n-1) + \gamma \sqrt{n-1} + \chi (n-1)^\alpha,
\label{Fden:SI}
\end{eqnarray}  
using the same notations as in the Ref.~\cite{Weitz2013} of the main text. The value of $\alpha$ has been thoroughly explored in this work and typically lies between 1.5 and 2 in the case of coupling to the membrane curvature. The exact value depending on the membrane surface tension, going from $\alpha=2$ at low tensions to lower values at biological tensions.  Note that an exponent $\alpha=2$ also corresponds to the case where the repulsion is pairwise additive with a range larger than the typical cluster diameter. Indeed, in this case $F_{\rm rep}(n) \propto n(n - 1)/2 \propto n^2$. 

\section{Numerical calculation of $g(n)$~-- resolution of Eq.~\eqref{desc:rec}}

We assume that for $n$ sufficiently large, $c_n$ is vanishingly small. We thus set $g(n_{\rm max})>0$ and $g(n_{\rm max}+1)=0$. After calculating $g(1)$ through the recurrence relation, which only depends on $c_1$ and $g(n_{\rm max})$, we find their respective values by solving numerically, with the help of the {\sc Mathematica} software, the two equations
\begin{eqnarray}
1 & = & g(1)  \\
\phi & = & \sum_{n=1}^{n_{\rm max}}n c_n^{(0)} \left[ \frac{c_1}{c_1^{(0)}} \right]^n g(n) ,
\end{eqnarray}
the latter equation deriving from $\sum n c_n = \phi$ through Eq.~\eqref{cnvscn0}. Thanks to the linear character of the recurrence, this is equivalent to setting $g(n_{\rm max})=1$ and solving the unique equation $\sum_{n=1}^{n_{\rm max}}n c_n^{(0)} \left( c_1/c_1^{(0)} \right)^n g(n) = g(1) \, \phi$ where $c_1$ only is unknown. Once $c_1$ is known, the whole sequence $(g(n))_{n\leq n_{\rm max}}$ can be calculated, from which the $c_n$ derive, as desired (see Fig.~\ref{gofn_rec}). We shall see below that the bimodal character of the cluster-size distribution is preserved out of equilibrium. In practice, $g(n$) is decreasing with $n$ and is essentially constant around the multimer peak and beyond. We denote by $g(\infty)$ its limit. 

\section{Taylor expansions in the Gaussian-peak approximation}

\begin{figure}[h]
\centering
    \includegraphics[scale=0.55]{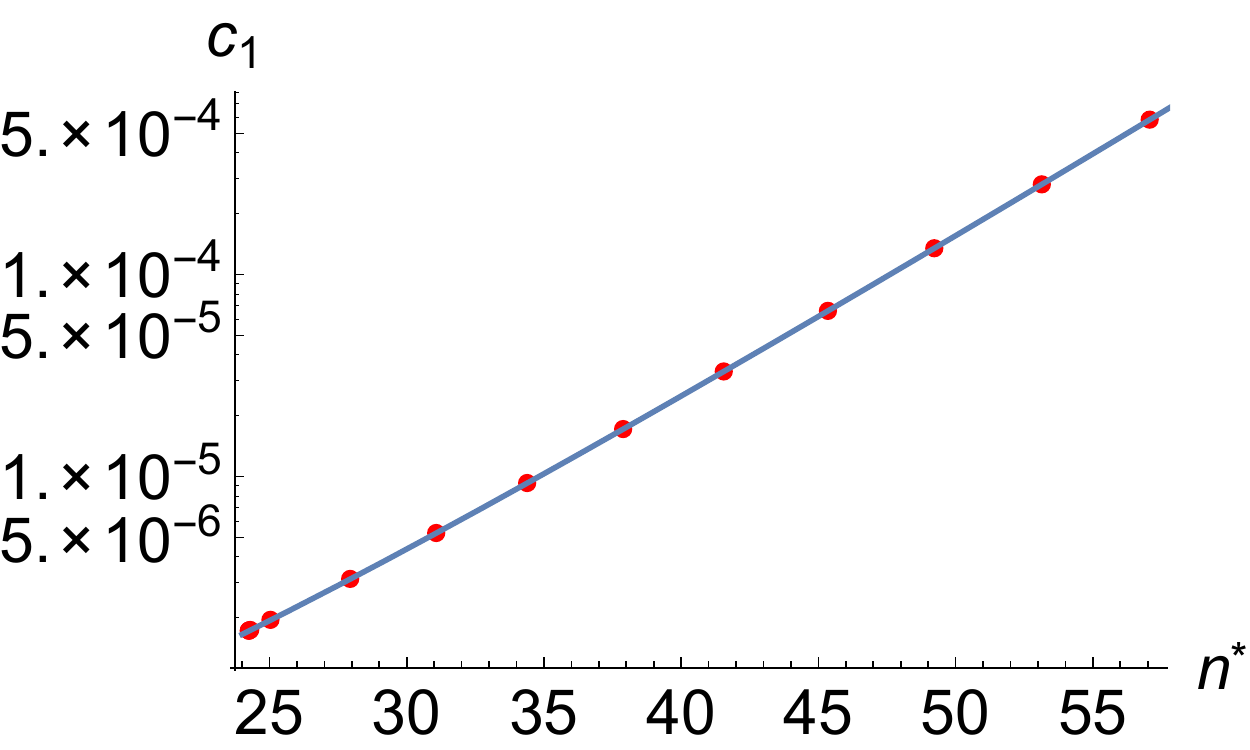} \caption{Comparison of the monomer concentration $c_1$ vs the multimer peak position $n^*$ found by calculation (Eq.~\eqref{c1BIS}, continuous line) and by the recurrence (dots). The parameters are the reference ones with $\phi=0.1$. Both $\sigma$ and $n^*$ depend on the recycling rate $j_\text{off}$ and the red dots from the left to the right correspond to the recycling rates $j_\text{off}=0, 10^{-11},10^{-10},10^{-9}, 10^{-8}, 10^{-7}, 10^{-6}, 10^{-5}, 10^{-4}, 10^{-3}$ and $10^{-2}$~s$^{-1}$. Lin\-ear-log coordinates. 
\label{figc1}} 
\end{figure}

\begin{figure}[h]
\centering
    \includegraphics[scale=0.55]{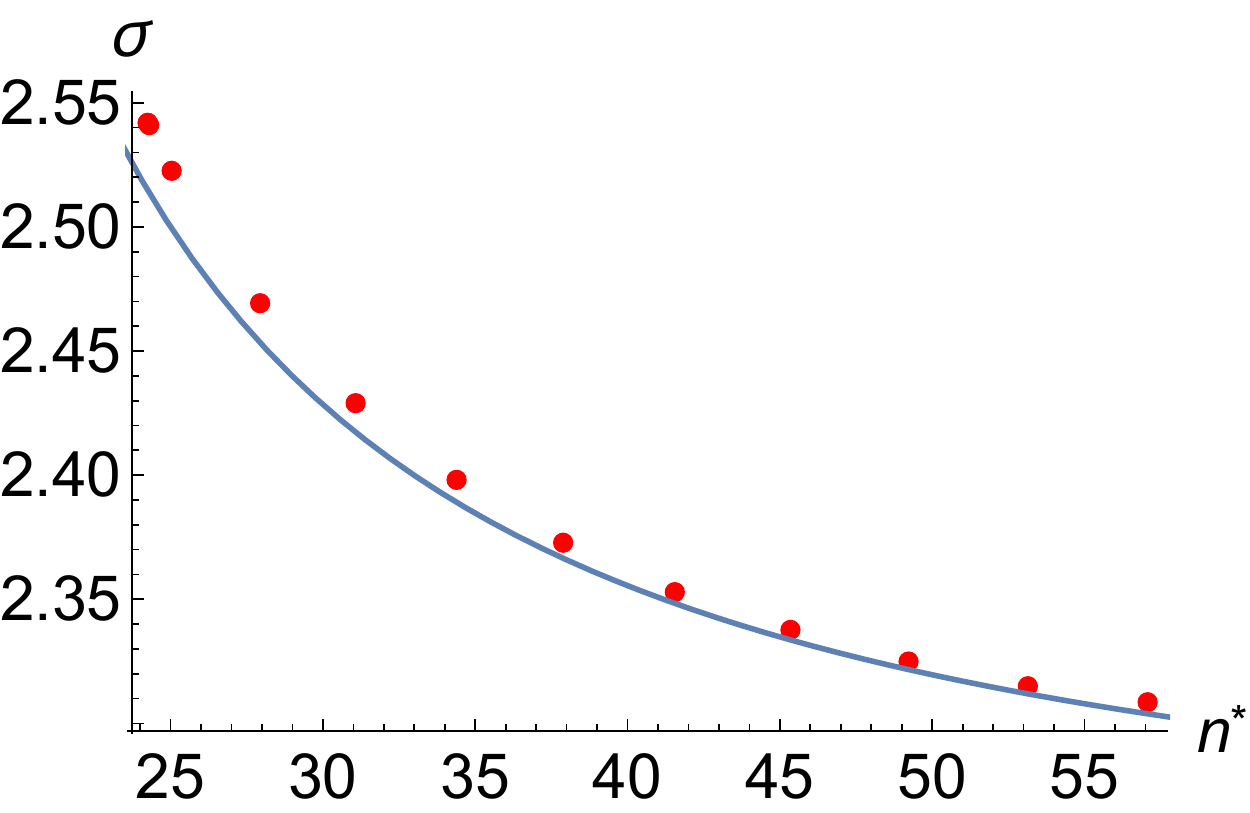} 
\caption{Standard deviation $\sigma$ of the Gaussian vs the multimer peak position $n^*$, found by analytical calculation (Eq.~{\bf [\,20\,]}, 
continuous line) and by fitting the numerical Gaussians (dots). The parameters are the reference ones with $\phi=0.1$. We have chosen the same recycling rates $j_\text{off}$ as in Fig.~\ref{figc1}. 
\label{figsigm}}
\end{figure}

\begin{figure}[h]
\centering
    \includegraphics[scale=0.55]{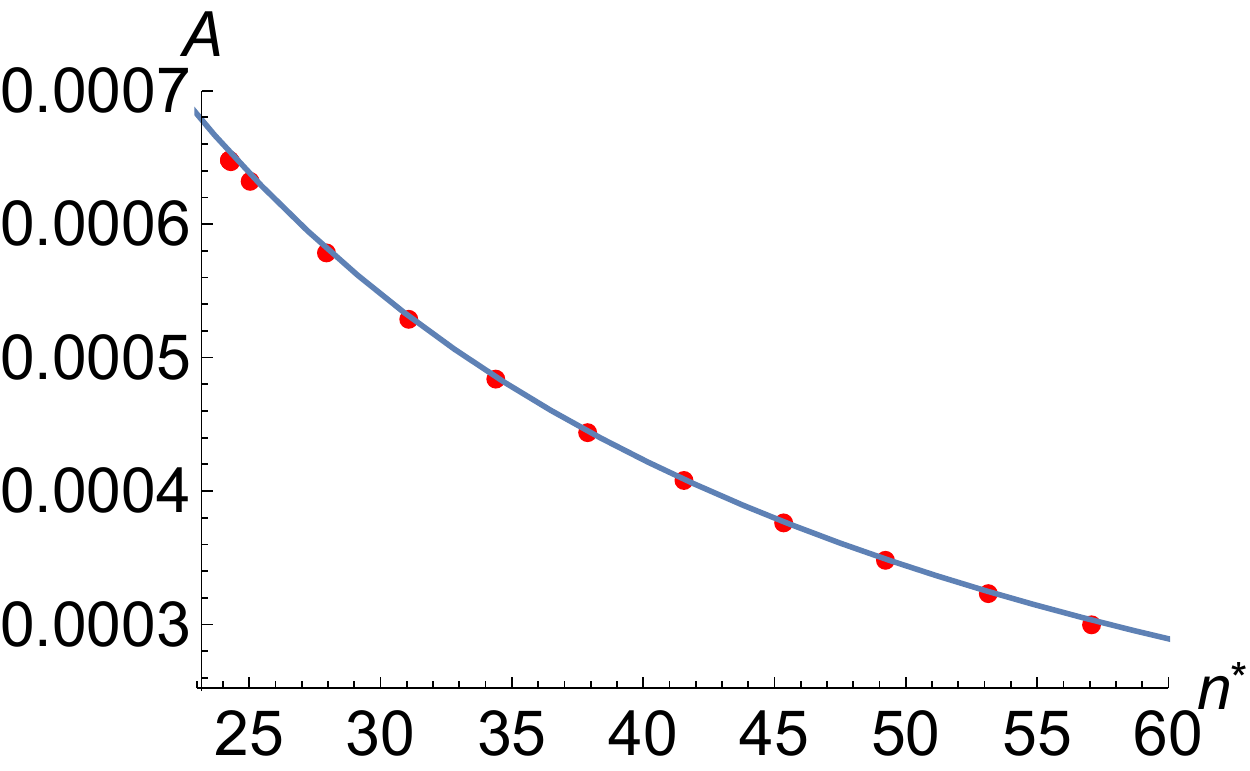} 
\caption{Comparison of the amplitude $A$ of the Gaussian vs the multimer peak position $n^*$ found by analytical calculation (lines) and by the recurrence (dots). The analytical curve comes from Eq.~\eqref{eqA}, having omitted $c_1 \ll \phi$ in the numerator. The parameters are again the reference ones with $\phi=0.1$. We have chosen the same recycling rates $j_\text{off}$ as in Fig.~\ref{figc1}. 
\label{A}}
\end{figure}

In this appendix, we derive Eqs.~\eqref{c1BIS}. The approximations used here are validated by the agreement between numerical and analytical calculations displayed in Figs.~\ref{figc1}, \ref{figsigm}, \ref{A}, and \ref{alpha=2}.

We start from Eq.~\eqref{master2}, itself deriving from the master equation~\eqref{master} for $n\geq2$ and the approximation of the distribution $(c_n)$ proposed in Eq.~\eqref{cn} and justified in the main text. We use several Taylor-expansions:
\begin{equation}
e^{-\frac{1\pm2\,\left(n-n^*\right)}{2\,\sigma^2}} = 1 -\frac1{2\,\sigma^2} \mp \frac{n-n^*}{\sigma^2} + \mathcal{O}[(n-n^*)^2]
\label{taylor:exp1}
\end{equation}

In addition, at order 1 in $(n-n^*)$,
\begin{eqnarray}
\sqrt{n}-\sqrt{n-1} & \simeq& \sqrt{n^*}-\sqrt{n^*-1}  \nonumber \\
 &  + & \left(  \frac{1}{2 \sqrt{n^*}} - \frac{1}{2 \sqrt{n^*-1}} \right) \left(n-n^*\right) \nonumber \\
& \simeq & \frac{1}{2 \sqrt{n^*}}+\frac{1}{8}\,\left(\frac{1}{n^*}\right)^{3/2} \nonumber \\
 & - &\frac{1}{4}\,\left(\frac{1}{n^*}\right)^{3/2}\,\left(n-n^*\right)
\label{sqrt}
\end{eqnarray}
where the coefficients of the expansion have been themselves expanded up to order 3/2 in powers of $1/n^*$. Similarly,
\begin{eqnarray}
\sqrt{n-1}-\sqrt{n-2} & \simeq& \frac{1}{2 \sqrt{n^*}}+\frac{3}{8}\,\left(\frac{1}{n^*}\right)^{3/2} \nonumber \\
 & - & \frac{1}{4}\,\left(\frac{1}{n^*}\right)^{3/2}\,\left(n-n^*\right)
\label{sqrt2}
\end{eqnarray}

Furthermore, for any $1<\alpha \leq 2$,
\begin{eqnarray}
n^\alpha-\left(n-1\right)^\alpha&\simeq &\alpha \, n^{*(\alpha-1)}-\frac{\alpha(\alpha-1)}{2}\, n^{*(\alpha-2)} \nonumber \\
 & + & \alpha(\alpha-1) n^{*(\alpha-2)} \,\left(n-n^*\right)
\label{alpha}
\end{eqnarray}
and 
\begin{eqnarray}
\left(n-1\right)^\alpha-\left(n-2\right)^\alpha&\simeq &\alpha \, n^{*(\alpha-1)}-\frac{3\alpha(\alpha-1)}{2}\, n^{*(\alpha-2)} \nonumber \\
 &  + & \alpha(\alpha-1) n^{*(\alpha-2)} \,\left(n-n^*\right)
\label{alpha2}
\end{eqnarray}
at order 1 in $\left(n-n^*\right)$ and order $(\alpha-2)$ in $n^{*}$. 

Plugging these expansions in Eq.~\eqref{master2}, several terms cancel and we get
\begin{eqnarray}
\hat j_{\rm off} & \simeq & e^{F'(n^*)}\left[- \frac1{2\sigma^2} - \frac{\gamma}{4(n^*)^{3/2}} \right. \nonumber \\
  & + & \left.  \frac{\alpha(\alpha-1)}{2} \chi n^{*(\alpha-2)} - \frac{n-n^*}{\sigma^2} \right] \nonumber \\
  & + & c_1 \left( - \frac1{2\sigma^2} + \frac{n-n^*}{\sigma^2} \right)
\label{ouf}
\end{eqnarray}
where we have used $ \gamma/(2\sqrt{n^*}) - f_0 + \chi \alpha \, n^{*(\alpha-1)} \simeq F'(n^*)$ at large $n^*$ (see Eq.~\eqref{Fden}). We have kept $F'(n^*)$ in the exponential because it does not tend to zero at large $n^*$ (or large $\sigma$), contrary to the other terms, which justifies to expand the exponentials. The relation~\eqref{ouf} holds for all $n$ close to $n^*$. Equating order-1 terms in $(n-n^*)$, we get $c_1=e^{F'(n^*)}$, i.e.  Eq.~\eqref{c1BIS}. Replacing $c_1$ in the above equation and equating order-0 terms, we get an expression for $1/\sigma^2$ leading in turn to
\begin{align}
\label{sigm}
\sigma \simeq \left[\alpha(\alpha-1) \chi \left( n^* \right)^{\alpha-2} - \frac{\gamma}4  \left( n^* \right)^{-3/2} -\frac{\hat j_\text{off}}{c_1(n^*)}  \right]^{-1/2}.
\end{align}
In the special case where $\alpha=2$ and the recycling $j_\text{off}$ is low, this relation becomes Eq.~{\bf [\,20\,]} 
in the main text.

Strictly speaking, to infer two equations from the relation in Eq.~\eqref{ouf} valid for any $n$ close to $n^*$, we only need two different values of $n$, e.g. $n^*$ and $n^*+1$. This justifies the validity of the conditions in Eq.~\eqref{conds} even though $\sigma$ is not very large. To finish with, note that no more than two relations can be found using the above expansion in powers of $(n-n^*)$, even though going to higher orders, because a Gaussian probability distribution and its moments are fully characterized by only two quantities, its expectation value $n^*$ and its standard deviation $\sigma$. For example, we have checked that going to the order 2, we recover the same relation as compared to order~0.

\section{Additional data and parameter sets}

\subsection{Graphs for the reference parameter set}

We provide some additional graphs showing the very good agreement between our analytical approach and numerical results.

\section{Additional parameter sets}

\begin{figure}[h]
\centering
\includegraphics[scale=0.55]{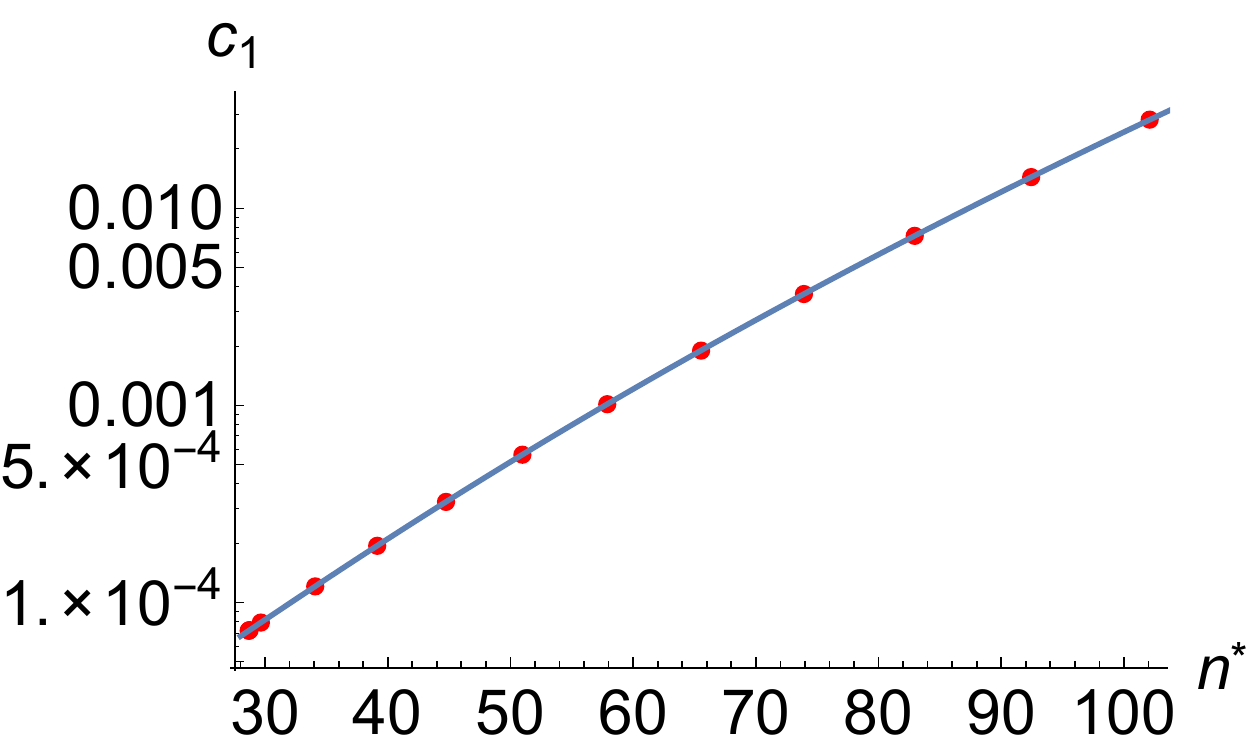} 
\includegraphics[scale=0.55]{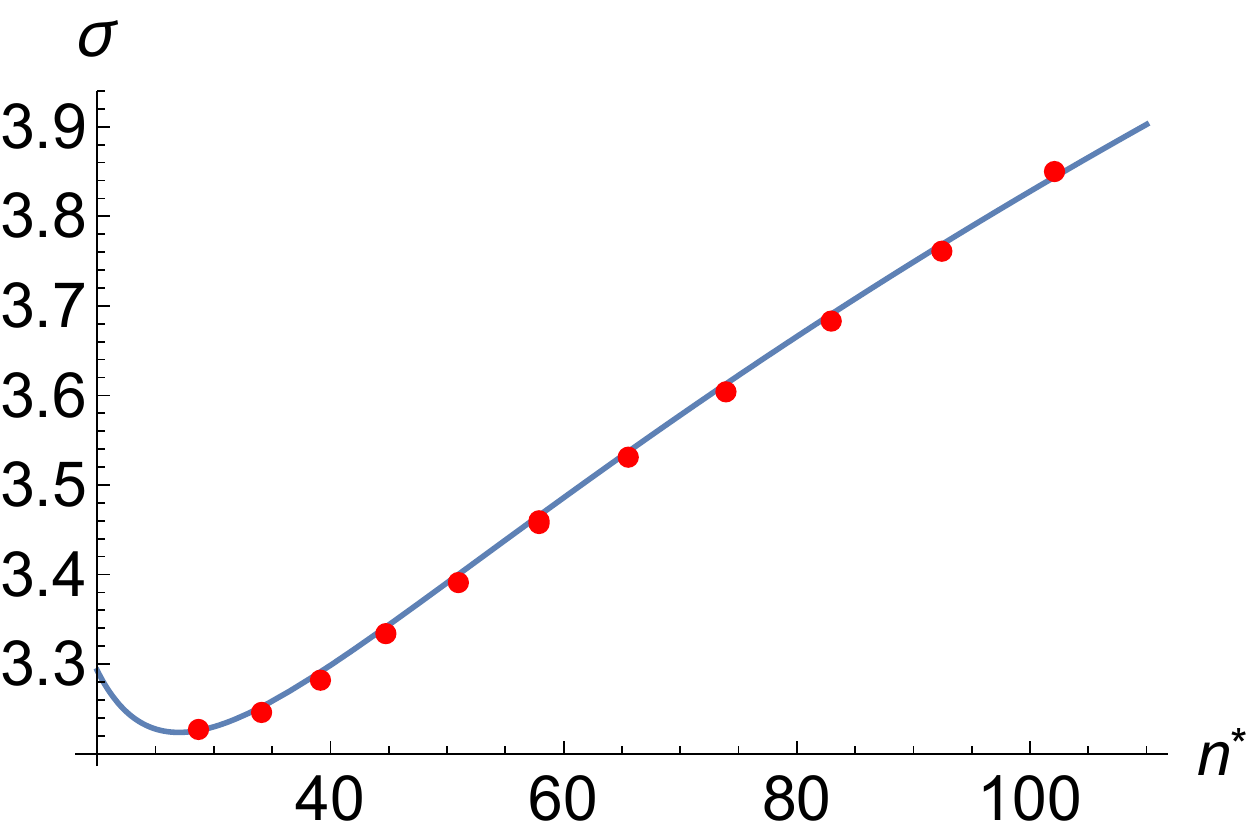} 
\includegraphics[scale=0.55]{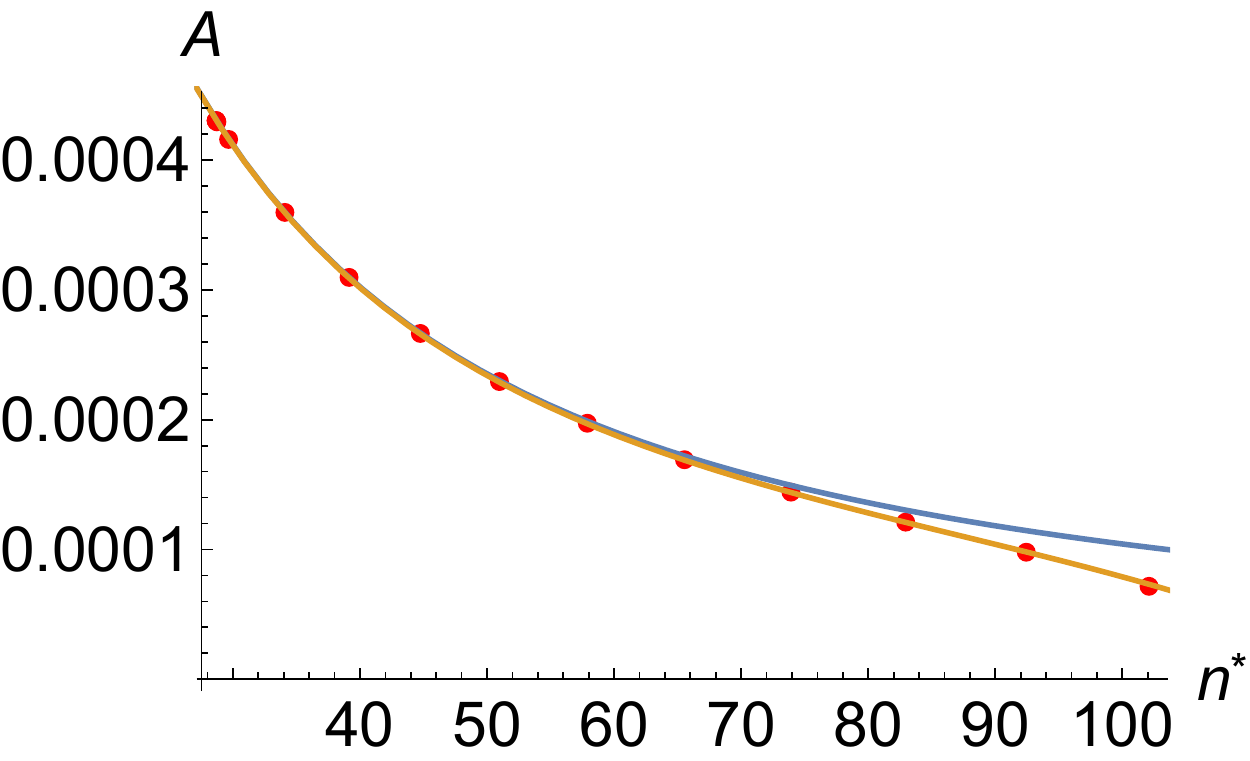} 
 \includegraphics[scale=0.55]{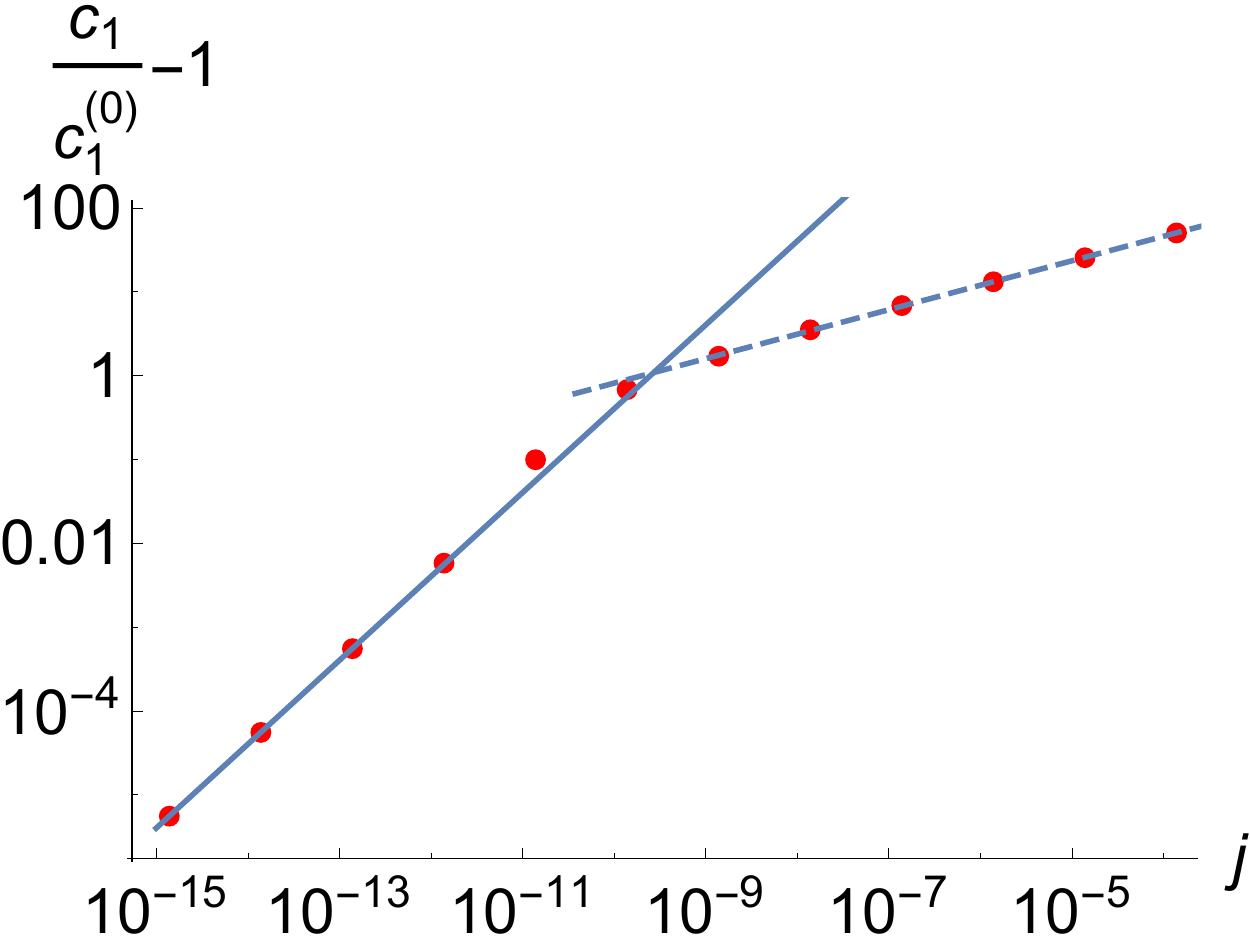}
\caption{Same as Figs.~\ref{figc1}, \ref{figsigm}, \ref{A} and Fig.~\ref{c1:joff} of the main text (right) with the parameter set $\alpha=3/2$, $f_0=20$, $\gamma=27$, $\chi=1$ and $\phi=0.1$. In the three top panels, $j_{\rm off}$ takes the values 0 and from $10^{-10}$ to 1~s$^{-1}$. Here $w_{\infty} \simeq 1.76 \times 10^{10}$, $c_1^{(1)} \simeq 2.86 \times 10^5$ and $\nu \simeq 0.29$. In the third panel, we have shown the analytical curves coming from Eq.~\eqref{eqA} without (blue curve) or with (orange curve) the term $c_1$ in the numerator.
\label{alpha=2}}
\end{figure}

In order to check the robustness of our results with respect to the values of the different parameters and the concentration $\phi$, we have tested different parameter sets compatible with experimental facts. The problem is analytically tractable for $\alpha=2$ only, because one ends with an hypergeometric ODE, an analytical solution of which is known. However, $\alpha$ probably lies between 1.5 and 2 (Ref~\cite{Weitz2013} of the main text), and we have therefore tackled numerically two parameter sets with $\alpha=3/2$. In Fig.~\ref{alpha=2}, we have plotted the so-obtained results for one such parameter set: $f_0=20$, $\gamma=27$, $\chi=1$ and $\phi=0.1$. Our qualitative conclusions remain unchanged, even though some details can be different, such as the monotonicity of $\sigma(n^*)$, owing to Eq.~{\bf [\,20\,]}. 
The exponent $\nu$ at intermediate recycling is measured to be $\nu\simeq0.29$ in this case.

We have also tested a parameter set initially designed to fit the cluster sizes of the Ref.~\cite{Destain08} of the main text, resulting from Monte Carlo simulations: $\alpha=3/2$, $f_0=19$, $\gamma=27.295$ and $\chi=0.6073$ (again in units of $k_BT$). We have also focussed on $\phi=0.1$. Again our qualitative conclusions remain identical (data not shown) even though the clusters are typically bigger ($n_0^* \simeq 47$ at equilibrium and $n^*\simeq 187$ if $j_{\rm off}=10^{-3}$; $n_{\rm max}=300$ is required in this case). The agreement between numerical and analytical calculations remains excellent. The  measured exponent $\nu$ at intermediate recycling is also $\nu\simeq0.29$.

To finish with, we have explored two different surface fractions for the reference parameter set ($\alpha=2$, $f_0=\gamma=20$ and $\chi=0.1$), namely $\phi=0.01$ and $\phi=0.3$. We recall that biological protein surface fractions are probably between 0.1 and 0.3. Our conclusions also remain identical on a qualitative level. In particular, the measured exponent $\nu$ at intermediate recycling remains equal to 0.32 for $\phi=0.3$ whereas it is slightly lower ($\nu \simeq 0.31$) for $\phi=0.01$.


\section{Calculation of $c_1^{(1)}$}

The recurrence~\eqref{wn:rec} cannot be solved exactly because it depends on the choice of the potential $G(n)$ through $\Gamma(n)$. Like $g(n)$, this recurrence relation appears to have a stable solution when solved in a descending way with the condition that $w_1=0$. Eq.~\eqref{cnvscn0} then yields
\begin{equation}
A=c_{n^*}^{(0)} \left[ \frac{c_1}{c_1^{(0)}} \right]^{n^*} g(n^*) \, ,
\label{cnvscn0s}
\end{equation}
when taken at $n=n^*$. We set $c_1=c_1^{(0)} + c_1^{(1)} \, j + \mathcal{O}(j^2)$. Since $c_1^{(0)}$ is known from equilibrium, it is the coefficient $c_1^{(1)}$ that we are looking for.  We also set $A=A_0 - A_1 \,  j
+ \mathcal{O}(j^2)$. Close to equilibrium, $n^*$ is given by Eq.~\eqref{nstarorder1}. After taking its logarithm and keeping the $\mathcal{O}(j)$ terms only, Eq.~\eqref{cnvscn0s} leads to
\begin{equation}
c_1^{(1)} = \frac{c_1^{(0)}}{n_0^*}\left(w_{n^*} - \frac{A_1}{A_0} \right) \, .
\end{equation}
Notably, we have used $\ln c_{n^*}^{(0)} \simeq \ln A_0 - (n^*-n^*_0)^2/(2\sigma_0^2) \simeq \ln A_0 - \sigma_0^2\ln^2 (c_1/c_1^{(0)})/2 = \ln A_0 + \mathcal{O}(j ^2)$ owing to the Gaussian character of $c_n$ close to $n_0^*$. In practice, $w_{n^*}$ appears to be very close to the limit $w_\infty$ of $(w_n)$ and this quantity is much larger than $A_1/A_0$, which leads to the simpler approximate relation $c_1^{(1)} \simeq c_1^{(0)} / n_0^* \, w_{\infty}$ used in the main text.

\section{Study of the sequence $g(n)$}

\subsection{Numerical example}

\begin{figure}[h]
\centering
\includegraphics[scale=0.55]{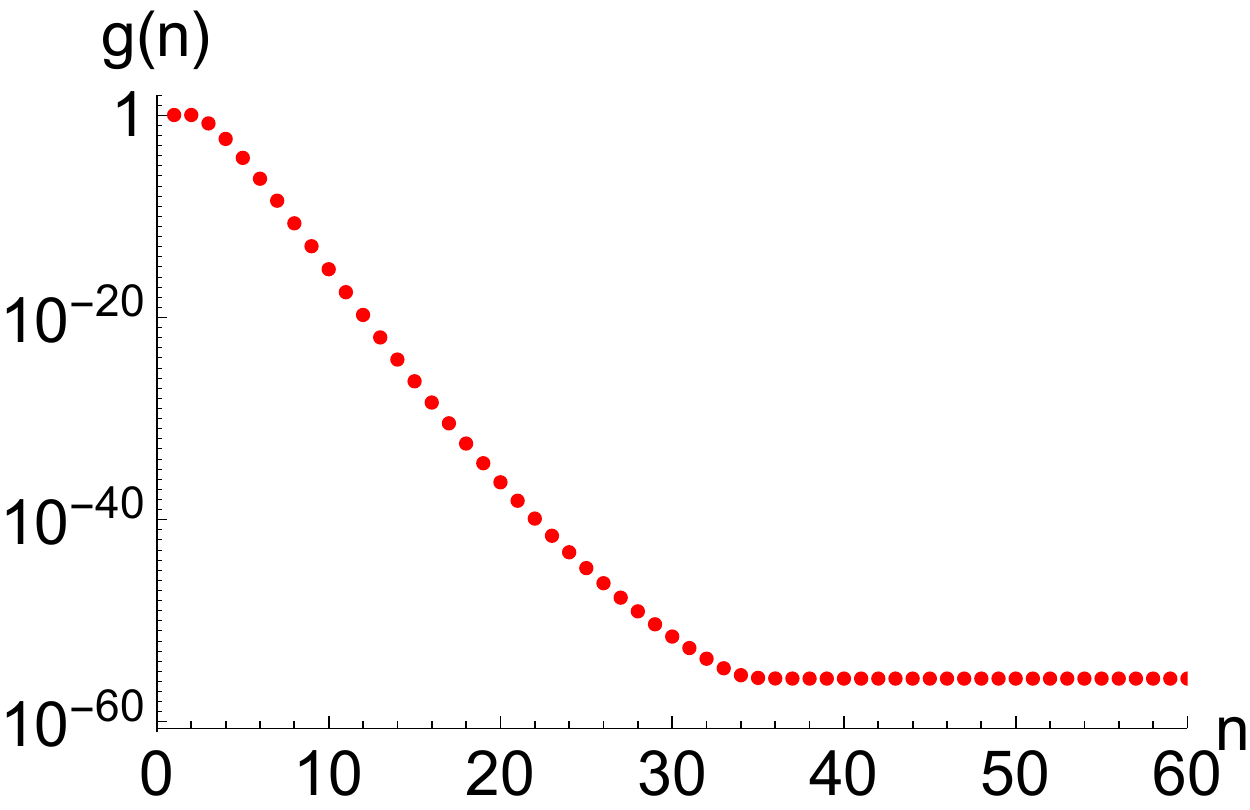} 
\caption{Numerical solution of the recurrence relation~\eqref{desc:rec} for the following parameter set: $\alpha=2$, $f_0=20$, $\gamma=20$, $\chi=0.1$, $\phi=0.1$ and $j_{\rm off}=10^{-5}$~s$^{-1}$. Linear-log coordinates.
\label{gofn_rec}}
\end{figure}

An example of numerical solution of the descending recurrence relation~\eqref{desc:rec} is given in Fig.~\ref{gofn_rec} for the reference parameter set. An exponential decay follows a short plateau near $n=1$, before a sudden saturation to $g(\infty)$. This typical behavior, observed for all the parameter sets explored in this work, is fully exploited in the analytical approach and it is also given an analytical explanation below in terms of hypergeometric functions. 

\subsection{Analytical study in the continuous limit}

Our goal here is to give an analytical solution to the recurrence relation~\eqref{desc:rec} on $g(n)$. We again consider it in the ascending way.

For any initial conditions $g(1)$ and $g(2)$, the space of solutions is a two-dimensional vector space $\mathcal{S}$. However, one of the eigenvalues associated with the recurrence relation is always smaller than 1 and the second one is close to 2. Therefore we expect that only a 1D sub-space $\mathcal{S}_1$ of the solutions of $\mathcal{S}$ do not diverge at large $n$, as expected from physical solutions of the recurrence relation. In order to characterize  $\mathcal{S}_1$, we consider the continuous limit of the recurrence relation, obtained by developing $g(n\pm1)$ up to order 2: $g(n\pm1) \simeq g(n) \pm g'(n) + \frac12 g''(n)$.

We thus get an order 2 ordinary differential equation (ODE) on $g$, and we shall select the solutions having a finite limit when $n \rightarrow \infty$.  Using $c_1/c_1^{(0)} =\Gamma(n^*)$ and $j = \hat j_{\rm off}/k'/c_1^{(0)}$, the ODE more precisely reads
\begin{equation}
\left[ \Gamma(n^*)- \Gamma(n)\right]g'(n) + \frac12 \left[ \Gamma(n^*)+ \Gamma(n)\right]g''(n)=jg(n).
\end{equation}
The second approximation will consist of keeping only the two leading terms in the expression of $G'(n)$ in Eq.~\eqref{def:Gamma}: $ G'(n) \simeq - (f_0+\mu) + \alpha \chi n^{\alpha-1}$. This is valid in the large $n$ limit where $1/\sqrt{n}$ goes to 0. In order to go further into the analytical investigation, we now set $\alpha=2$ because an analytical solution of the ODE exists in this case. We recall that  $\alpha=2$ is of physical relevance for a bio-membrane in the low-tension limit (see Ref~\cite{Weitz2013} of the main text). We have previously explored numerically the value $\alpha = 3/2$ in this SI. 

Defining $\Gamma_0=e^{-(f_0+\mu)}$ and the new variable $t=e^{2\chi n}\in \mathbb{R}^{+*}$ we obtain the new ODE 
\begin{align}
2\chi \Gamma_0 \left[ t^*(\chi+1)+t(\chi-1)Ê\right] \, t \, g'(t) &  \nonumber \\
 + \; 2 \chi^2 \Gamma_0 (t^*+t)\, t^2 \, g''(t) & =  j g(t).
\label{zeODE} 
\end{align}
with $t^*=e^{2\chi n^*}$, which can in turn be written as 
\begin{align}
 t^2 (t+t^*) \, g''(t)  + a \, t (t+t^* u_1) \, g'(t)  =  b t^* g(t),
\end{align}
by setting $a=1-1/\chi$, 
\begin{equation}
\label{def:b}
b=\frac{j}{2 t^* \chi^2 \Gamma_0}=\frac{j_{\rm off}}{2 c_1 k' \chi^2},
\end{equation}
and $u_1=(\chi+1)/(\chi-1)$. If we define the new variable $u=t/t^*=e^{2\chi (n-n^*)}\in \mathbb{R}^{+*}$, this ODE simplifies to
\begin{align}
u^2 (u+1) \, g''(u)  + a \, u (u+ u_1) \, g'(u)  =  b g(u).
\end{align}
It can be shown to be equivalent to a hypergeometric equation~\cite{Abramowitch} by setting $g(u)=u^\lambda \,  h(u)$ and suitably choosing $\lambda$, as discussed now.

\subsection{Solution of the hypergeometric equation}

Setting $z=-u \in \mathbb{R}^{-*}$, $h(u)=y(z)$, and choosing 
\begin{align}
\lambda=\lambda_\pm = ( 1 -a \, u_1 \pm \sqrt{(a \, u_1-1)^2+4 b})/2,
\label{lambdas}
\end{align}
the previous ODE becomes the hypergeometric equation in its usual form:
\begin{align}
z (1-z) \, y''(z)  + [ 2(1-z)\lambda_\pm+a(u_1-z)] \, y'(z)  & \\
- [b+a(1-u_1)\lambda_\pm ] \, y(z) & =0. \nonumber
\end{align}
The two independent solutions of ODE are then given by hypergeometric functions~\cite{Abramowitch} (expressed in the variable $u$):
\begin{align}   
   g_\pm(u) =  
   u^{\lambda_\pm}  {_2F_1}\left(\lambda_\pm,\lambda_\pm +a-1;2\lambda_\pm+ a \, u_1 ;-u\right).
\end{align}
In the present context, $a\, u_1 - 1 = 1/\chi=1-a$ and $b \ll 1/\chi^2$ so that $\lambda_- \simeq -1/\chi$ and $\lambda_+ \simeq j/(2\chi \Gamma_0 t^*) \ll \lambda_-$. Hence
\begin{align}   
   g_-(u) \simeq 
   u^{-1/\chi} \,  {_2F_1}\left(-1/\chi,-2/\chi;1-1/\chi ;-u\right) 
\end{align}
at the lowest order in $j$ and                        
\medskip
\begin{align}   
   g_+(u) \simeq 
   u^{j/(2\chi \Gamma_0 t^*)}  {_2F_1}\left( j/(2\chi \Gamma_0 t^*),-1/\chi; 1+1/\chi ;-u\right).
\end{align}
Any solution $g$ can be written as a linear combination of $g_-$ and $g_+$. We shall now see that the requirement that $g(u)$ does not diverge when $u \rightarrow \infty$ restricts the accessible solutions. The additional condition $g(1)=1$ will set the unique solution to our initial problem.

\subsection{Asymptotic behaviors}

\begin{figure}[h]
\centering
\includegraphics[scale=0.55]{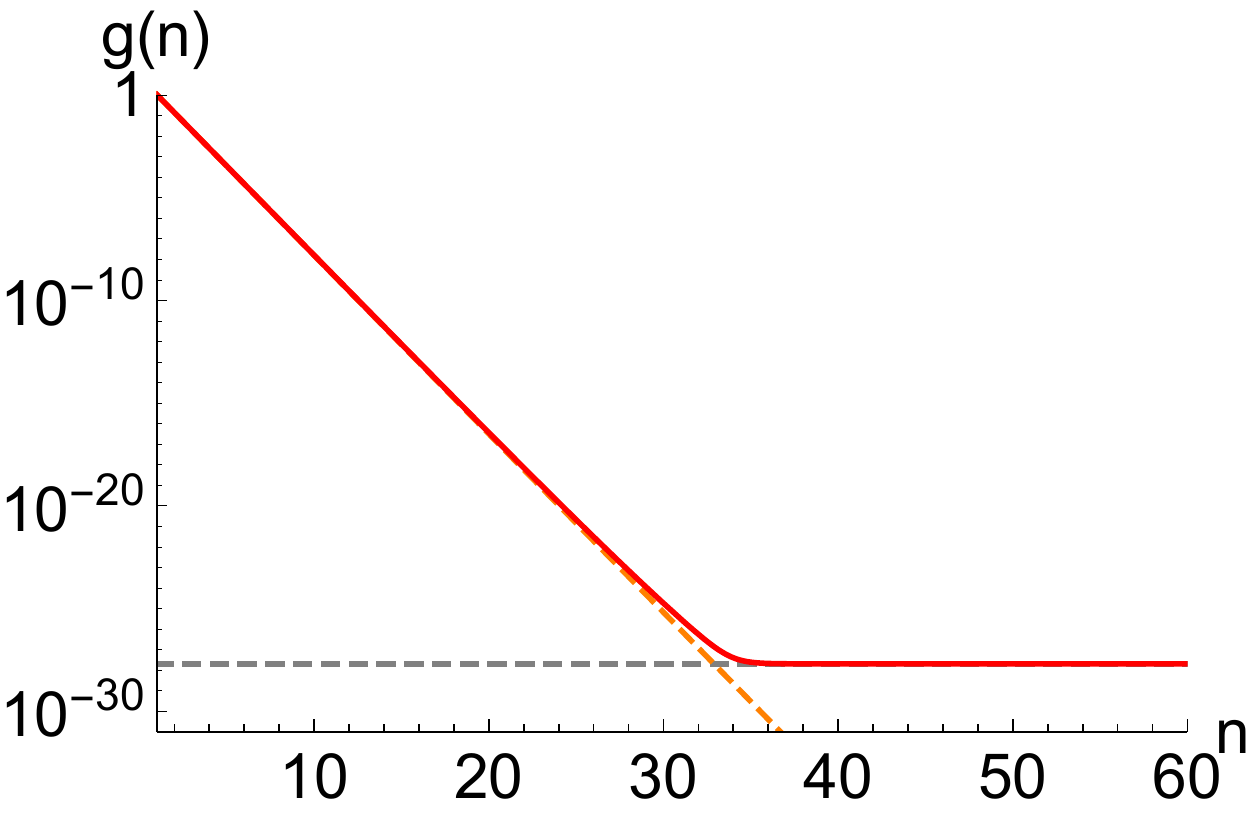} 
\caption{Analytic solution of the hypergeometric ODE~\eqref{zeODE} for the same set of parameters as in Fig.~\ref{gofn_rec}. The full red line is the solution given by Eq.~\eqref{def:u}, whereas the orange and grey dashed lines are the small and large $n$ approximations given by Eqs.~\eqref{g:small:n} and \eqref{g:large:n}, respectively. The crossover between both regimes, defined as the intersection between both approximations, occurs at $n=\tilde n<n^*$. Linear-log coordinates.
\label{gofn_edo}}
\end{figure}

Given that $z\in \mathbb{R}^{-*}$, we can use the relation~(\cite{Abramowitch}, Eq.~15.3.7):
\begin{align}
&z^A{_2F_1}(A,B;C;z)  \nonumber \\
=&\frac{\hat \Gamma(C)\hat \Gamma(B-A)}{\hat \Gamma(B)\hat \Gamma(C-A)} {_2F_1}\left(A,1-C+A;1-B+A;\frac1z\right) \nonumber \\
+&z^{A-B}\frac{\hat \Gamma(C)\hat \Gamma(A-B)}{\hat \Gamma(A)\hat \Gamma(C-B)} {_2F_1}\left(B,1-C+B;1-A+B;\frac1z\right). 
\end{align}
Here $\hat \Gamma(z)$ denotes Euler's Gamma function~\cite{Abramowitch} (not to be confused with $\Gamma(n)$ as defined in Eq.~\eqref{def:Gamma}). The divergence at $-\infty$ comes from the second term of the r.h.s. because in our case  $A-B=\lambda_\pm-(\lambda_\pm+a-1)=1-a=1/\chi>0$. The hypergeometric function appearing in this second term appears to be identical for both functions $g_+$ and $g_-$ [and equal to $_2F_1(\lambda_+ - 1/\chi,\lambda_- - 1/\chi;a;1/z)$]. Defining 
\begin{align}
g(u) = p_- g_-(u) - p_+ g_+(u)
\label{def:u}
\end{align}
with 
\begin{eqnarray}
p_-&=&\hat \Gamma (\lambda_-) \hat \Gamma (2\lambda_++1+1/\chi) \hat \Gamma (\lambda_-+1+2/\chi), \nonumber  \\
p_+&=&\hat \Gamma (\lambda_+) \hat \Gamma (2\lambda_-+1+1/\chi) \hat \Gamma (\lambda_++1+2/\chi),  \nonumber \\
\end{eqnarray}
divergences cancel and we obtain a solution of our original ODE that does not diverge at infinity, as requested. This solution is represented (after normalization to fulfill $g(1)=1$) in function of the original variable $n$ in Fig.~\ref{gofn_edo}, with the same parameters as in Fig.~\ref{gofn_rec}. We observe an exponential decay at small $n$ and a saturation at large $n$, as in the numerical solution of Fig.~\ref{gofn_rec}. The crossover between both regimes appears at a value $\tilde n$ that agrees remarkably well between the numerical and analytical solutions, as expected from the fact that the analytical approximation becomes exact in the large $n$ limit. By contrast, the fall between $n=1$ and $\tilde n$ is clearly underestimated by the analytical approach because we have neglected the $\sqrt{n}$-term in $G(n)$ when approximating $G'(n)$ above. We shall return to this issue below. 

From Eq.~\eqref{def:u} and the behavior of hypergeometric functions close to 0 and $-\infty$~\cite{Abramowitch}, we obtain
\begin{equation}
g(n) \simeq p_- \, e^{2\chi \lambda_- (n-n^*)} \simeq p_- \, e^{-2 (n-n^*)}
\label{g:small:n}
\end{equation} 
for small values of $n$ and
\begin{eqnarray}
 g(n) & \longrightarrow & \ell(\lambda_-,\lambda_+,\chi)  \nonumber \\
  &\equiv & \hat \Gamma(-\frac1{\chi}) \hat \Gamma (2\lambda_- +1+\frac1{\chi})\hat \Gamma (2\lambda_+ +1+\frac1{\chi})  \nonumber \\
& \times & \bigg[\frac{\hat \Gamma(\lambda_-)\hat \Gamma(\lambda_-+1+\frac2{\chi})}
{\hat \Gamma(\lambda_- -\frac1{\chi})\hat \Gamma(\lambda_- +1 + \frac1{\chi})}    \nonumber \\
&   &- \, \frac{\hat \Gamma(\lambda_+)\hat \Gamma(\lambda_++1+\frac2{\chi}) }
 {\hat \Gamma(\lambda_+ -\frac1{\chi})\hat \Gamma(\lambda_+ +1 + \frac1{\chi})} \bigg]
\label{g:large:n}
\end{eqnarray} 
at large $n$. Finally, this function has to be divided by $g(1)$ to fulfill the condition $g(1)=1$, as displayed in Fig.~\ref{gofn_edo}. Note that the previous expression is not defined when $1/\chi$ is an integer because $\hat \Gamma(-1/{\chi})$ diverges. However, $g(n)$ has a finite limit in this case owing to the properties of the function $\hat \Gamma$, so this is not a true singularity. 

From these asymptotic behaviors, we obtain the value $\tilde n$ of $n$ at the crossover between both regimes:
\begin{equation}
\tilde n \simeq n^* - \frac12 \ln \left| \frac{\ell}{p_-} \right|, 
\label{B17}
\end{equation}
by equating Eqs.~\eqref{g:small:n} and \eqref{g:large:n}. 

\subsection{Small values of $n$ ($n<\tilde n$)}

We have seen that the ODE fails to characterize precisely the function $g(n)$ on the interval $[1,\tilde n]$  because we have neglected the $\sqrt{n}$-term in $G(n)$ when approximating $G'(n)$. We now propose a different approach on this interval.

\begin{figure}[h]
\centering
\includegraphics[scale=0.55]{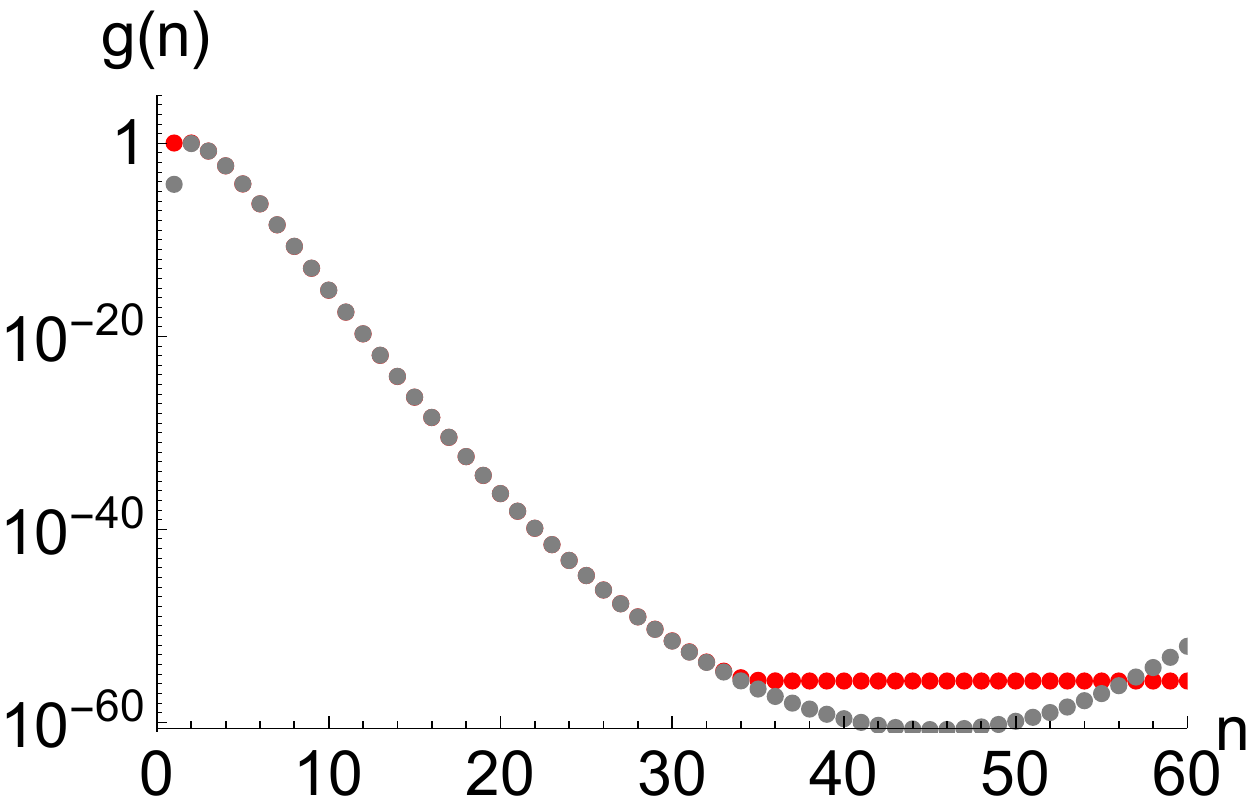} 
\caption{Comparison of the numerical solution of the recurrence relation~\eqref{desc:rec} as in Fig.~\ref{gofn_rec} (red dots) and of the approximate solution $g(n)=\kappa \, \left[c_1^{(0)}/c1 \right]^n e^{G(n)}$ with $\kappa \simeq 2\times 10^{-3}$ a suitably chosen constant (gray dots). The dots are almost superimposed for $2 \leq n \leq \tilde n$. Linear-log coordinates.
\label{gofn_approx}}
\end{figure}

We start from Eq.~\eqref{desc:rec}, rewritten as
\begin{equation}
\Gamma(n)[g(n-1)-g(n)] = \Gamma(n^*) [g(n)-g(n+1)] 
\end{equation}
where we have neglected $\hat j_{\rm off}$ because it is always much smaller than $c_1$. We now show that $g(n)=\kappa \, e^{G(n)} \Gamma(n^*)^{-n}$, with $\kappa$ a constant, is an approximate solution of this recurrence relation, as illustrated in Fig.~\ref{gofn_approx}. Indeed, replacing $g(n)$ by this expression in the equation above, we get
\begin{equation}
\Gamma(n)\left[e^{G(n-1)}-\frac{e^{G(n)}}{\Gamma(n^*) }\right] = \left[e^{G(n)}-\frac{e^{G(n+1)}}{\Gamma(n^*) }\right].
\end{equation}
This equality is verified because $\Gamma(n)e^{G(n-1)} \equiv e^{G(n)}$ and $\Gamma(n)e^{G(n)} \simeq e^{G(n+1)}/e^{G''(n)}$ with $e^{G''(n)} \simeq e^{2\chi} \simeq 1$ if $\chi \ll 1$.

In conclusion, we have argued so far that when $n<\tilde n$, $g(n)=\kappa \, \left[c_1^{(0)}/c_1 \right]^n e^{G(n)}= \kappa \, e^{G(n)} \Gamma(n^*)^{-n}$ and when $n>\tilde n$, 
$g(n) \simeq g(\tilde n) =  \kappa \, e^{G(\tilde n)} \Gamma(n^*)^{-\tilde n} = g(\infty)$ is almost constant. Note that the recycling rate $j_{\rm off}$ does not play significantly on the value of $g(n)$ for $n < \tilde n$, but it plays on the value of $\tilde n$ and consequently on the value of $g(\infty)$. 

\subsection{Study of $\tilde n(b)$}

The crossover occurs at $n=\tilde n$, a function of the different parameters through Eq.~\eqref{B17}. In particular, it is a function of $j$ and $n^*$ through the ratio $b=j/(2 t^* \chi^2 \Gamma_0)$, i.e. through the ratio $j/e^{2\chi \, n^*}$. We now study the function $\tilde n(b)$ (all other parameters being fixed) in order to understand the observed scalings of $c_1(j)$, as displayed in Fig.~\ref{c1:joff} (right) for example. 

More precisely, $\tilde n$ depends on $b$ \emph{via} $\lambda_-$ and $\lambda_+$ (see Eq.~\eqref{lambdas}):
\begin{align}
\lambda_\pm = \frac1{2\chi}\left( -1 \pm \sqrt{1+4 \chi^2 \, b} \right),
\label{lambdasBIS}
\end{align}

We first study the function $f(b)= b \left|\ell(b)/p_-(b) \right|$. After a tedious but straightforward calculation, one shows that $f(b)$ has a finite limit (a function of $\chi$ only) when $b\rightarrow0$. It ensues that $\ln \left| \ell/p_- \right|=C(\chi)-\ln b$ at small $b$, with
\begin{equation}
C(\chi) = \ln \left[\frac{\sin \left( \frac{\pi}{\chi} \right)}{\chi \sin \left( \frac{2\pi}{\chi} \right) B \left( \frac{\pi}{\chi},-\frac{2\pi}{\chi} \right)}\right],
\end{equation}
where $B$ is Euler's Beta function~\cite{Abramowitch}. $C(\chi)$ has a finite value even if $1/\chi$ is an integer.

The above behavior of $\ln \left| \ell/p_- \right|$ has been established in the $b\rightarrow 0$ limit. In practice it appears to remain valid up to $b\sim1$ for the reference parameter set used so far, that is to say for all the values of $n^*$ and $j$ of interest in this work. 

Using Eqs.~\eqref{def:b} and \eqref{B17} one gets
\begin{align}
\tilde n = \tilde n_1 + (1-\chi) n^* + \frac12 \ln j,
\label{ntilde:nstar:j}
\end{align}
where 
\begin{align}
\tilde n_1 = \frac{f_0+\mu}2 -\frac12[C(\chi)+\ln 2] - \ln \chi
\end{align}
is independent of $n^*$ and $j$. Eq.~\eqref{ntilde:nstar:j} is used in the main text. Its validity can also be tested with the help of our numerical solution for $g(n)$. If $\tilde n$ is measured as the position of the maximum of $g''(n)$, then fitting the numerical values gives 0.88 and 0.59 instead of $1-\chi=0.9$ and $1/2$ for the reference parameter set. This is quite satisfactory and again validates our approximations.

In addition, for our reference parameter set and $\phi=0.1$, one gets $\tilde n_1\simeq -3.0$. Whenever $\tilde n_1 <0$ and $\ln j <0$, Eq.~\eqref{ntilde:nstar:j} implies that $\tilde n < n^*$. It follows that 
\begin{equation}
g(n^*)\simeq g(\tilde n) = \kappa \, e^{G(\tilde n)} \Gamma(n^*)^{-\tilde n}.
\label{last:rel}
\end{equation}

\section{Calculation of the exponent $\nu$}

We start from Eq.~\eqref{K:eq}:
\begin{equation}
K \simeq {n^*} \left[ \Gamma(n^*) \right]^{n^*-\tilde n} e^{G(\tilde n)-G(n^*)}
\end{equation}
with $K= (\phi \sqrt{\chi})/(\sqrt{\pi} \kappa \, c_{1}^{(0)})$. Taking the logarithm of this relation, we obtain
\begin{equation}
\ln \frac{K}{n^*} =  (n^* -\tilde n) \ln \Gamma(n^*) + G(\tilde n) - G(n^*).
\end{equation}
Owing to Eq.~\eqref{def:Gamma} and to $G(\tilde n) - G(n^*) \simeq  G'(n^*) (\tilde n-n^*) + 1/2 \, G''(n^*) (\tilde n-n^*)^2$,
we obtain the simple relation
\begin{equation}
\ln \frac{K}{n^*} \simeq  \frac12  G'' (n^*) (\tilde n-n^*)^2\, .
\end{equation}
Using $\tilde n \simeq \tilde n_1 + (1-\chi) n^* + \frac12 \ln j$ and the definition of $G$, 
\begin{equation}
\ln \frac{K}{n^*} \simeq  \left(\chi -\frac{\gamma}8 \frac1{(n^*)^{3/2}}   \right) \left(\tilde n_1-\chi n^* +\frac12 \ln j \right)^2.
\end{equation}
Since $j<1$ in the cases of interest, $\ln j <0$ and the expression in the square is negative. Thus
\begin{equation}
\frac12 \ln j \simeq -\tilde n_1+\chi n^* - \sqrt{\frac{\ln \frac{K}{n^*}}{\chi -\frac{\gamma}8 \frac1{(n^*)^{3/2}} }}.
\end{equation}
Approximating $n^*$ by $n_0^*$ in the slowly varying logarithm and expanding the square-root around $n^*_0$, we get 
\begin{eqnarray}
\frac12 \ln j & \simeq & {\rm Const.} +\chi n^* \nonumber \\
 &+ & \frac{3\gamma}{32} \sqrt{\ln \frac{K}{n^*_0}} \frac1{(n_0^*)^{1/4}} \frac{(n^*-n^*_0) }{\left[ \chi (n_0^*)^{3/2} -\frac{\gamma}8 \right]^{3/2}} 
\end{eqnarray}

This eventually yields
\begin{equation}
n^* \simeq {\rm Const.'} +  \frac{\nu}{2 \chi}  \ln j
\label{log:scalingBIS}
\end{equation}
with 

\begin{equation}
\nu = \left[ 1 + \frac{3\gamma}{32\chi} \sqrt{\ln \frac{K}{n^*_0}} \frac1{(n_0^*)^{1/4}\left[ \chi (n_0^*)^{3/2} -\frac{\gamma}8 \right]^{3/2}} \right]^{-1} .
\label{nu}
\end{equation}

\end{article}

\end{document}